\documentclass{aa}  
\usepackage{graphicx}
\usepackage{float}
\usepackage{booktabs}
\usepackage{placeins}
\usepackage[font=footnotesize,labelfont=bf]{caption}
\usepackage{subcaption}
\usepackage{tocloft}
\usepackage{xcolor}
\usepackage{colortbl}
\usepackage{arydshln}
\usepackage{multirow}
\usepackage{txfonts}
\usepackage[colorlinks=true, pdfborder={0 0 0}, linkcolor=black, citecolor=blue, urlcolor=blue]{hyperref}
\usepackage{longtable}
\RequirePackage{etex} 
\begin{document}

   \title{Heavy element enriched atmospheres and where they are born}

   \author{Barry O'Donovan
          \inst{1,2}
          \and
          Bertram Bitsch \inst{1}
          }

   \institute{Department of Physics, University College Cork, Cork, Ireland
                \and
                School of Physics, Trinity College Dublin, University of Dublin,
                Dublin 2, Ireland
             }

   \date{Received 18 August 2025; accepted 3 December 2025}
 
  \abstract
  {The heavy element content of giant exoplanets, inferred from structure models based on their radius and mass, often exceeds predictions based on classical core accretion. Pebble drift, coupled with volatile evaporation, has been proposed as a possible remedy to this with the level of heavy element enrichment a planet can accrete, as well as its atmospheric composition, being strongly dependent on where in the disc it is forming. We use a planet formation model which simulates the evolution of the protoplanetary disc, accounting for pebble growth, drift and evaporation, and the formation of planets from pebble and gas accretion. The growth and migration of planetary embryos is simulated in 10 different protoplanetary discs which have their chemical compositions matched to the host stars of the planets which we aim to reproduce, providing a more realistic model of their growth than previous studies. The heavy element content of giant exoplanets is used to infer their formation location and thus make a prediction of their atmospheric abundances. We focus here on giants more massive than Saturn, as we expect that their heavy element content is dominated by their envelope rather than their core. The heavy element content of 9 out of the 10 planets simulated is successfully matched to their observed values. Our simulations predict formation in the inner disc regions, where the majority of the volatiles have already evaporated and can thus be accreted onto the planet via the gas. As the majority of the planetary heavy element content originates from water vapor accretion, our simulations predict a high atmospheric O/H ratio in combination with a low atmospheric C/O ratio, in general agreement with observations. For certain planets, namely WASP-84b, these properties may be observable in the near future, offering a method of testing the constraints made on the planet's formation.}

   {}

   \keywords{giant planet formation --
                heavy element content --
                volatile content -- carbon-to-oxygen ratio
               }

   \maketitle

\section{Introduction}

The past two decades have seen an explosion in exoplanet discoveries, revealing an extraordinary diversity of planetary systems unlike our own \citep{Winn2015}. Among these, gas giants remain critical to our understanding of planet formation. Their prevalence, wide range of densities, and large radii make them both accessible to observation and influential in shaping planetary system architectures. Their formation not only affects the distribution and availability of materials for other forming planets (e.g. \citealt{Lambrechts2014,Bitsch2018}) but also influences orbital dynamics and stability within planetary systems (e.g. \citealt{BeaugeNesvorny2012, Schlecker2021, BitschIzidoro2023}) with important consequences for the presence of habitable planets (e.g. \citealt{Georgakarakos2018, Agnew2019}).

Classically, gas giant formation is understood via the core accretion paradigm: dust grains coagulate into pebbles and planetesimals, form a solid core ($\sim 10 M_\oplus$), which then accretes gas from the protoplanetary disk once that threshold is reached (e.g. \citealt{Pollack1996,Mordasini2012}). The pebble‐accretion variant enables rapid core growth via aerodynamic drag of mm–cm sized particles (e.g. \citealt{Ormel2010, LambrechtsJohansen2012}). Throughout growth, forming planets migrate via disk–planet torques (Type I) and eventually open gaps (Type II), moving them across diverse disk regions that differ in temperature and composition (e.g. \citealt{Oberg2011, Madhusudhan2014, Turrini2021, Bitsch2022, Penzlin2024}).

A major challenge to these models arises from the observations of \citet{Thorngren2016}, who analyzed 47 transiting giant exoplanets and derived their heavy‐element masses ($M_Z$) using interior structure models (later also confirmed by \citealt{Bloot2023}). \citet{Thorngren2016} found that many giant planets must host over 50-100$M_\oplus$ in heavy elements, indicating that many giant exoplanets contain significantly higher masses of heavy elements compared to the giant planets in our own Solar System, such as Jupiter and Saturn \citep{Vazan2018}. These large heavy element contents are also confirmed by more advanced interior models that take different internal structures into account \citep{Peerani2025}. These heavy element measurements are often too large to be confined solely to the core and implying heavy element enrichment of the gaseous envelope itself. Moreover, this enrichment shows only a weak correlation with host‐star [Fe/H], in contrast to earlier expectations \citep{Teske2019}. This poses a puzzle: classical formation pathways struggle to explain giant planets that are so largely enriched with heavy elements, especially when the host star lacks extreme metallicity \citep{Venturini2020}. A proposed solution to this problem could be found in collisions between growing planets that strip their envelope but retain the heavy element masses of the cores (e.g. \citealt{Ogihara2021}). However, the remaining systems in these models still contain multiple planets, whereas hot Jupiters, as used in the sample of \citet{Thorngren2016}, are mostly single planets (e.g. \citealt{Dawson2018}).

Pebble drift and evaporation at volatile evaporation fronts offers a solution to this puzzle by allowing heavy elements to be transported efficiently through the disc and be accreted into the atmospheres of growing planets.  As pebbles drift inward and evaporate at volatile evaporation fronts, they enrich the surrounding gas with heavy elements, which can then be accreted by forming gas giants (e.g. \citealt{Booth2017b, SchneiderBitsch2021a}). Models that account for this process yield results that are in slightly better agreement with \citet{Thorngren2016} compared to the classical core accretion scenario (e.g. \citealt{SchneiderBitsch2021a, Morbidelli2023, Bitsch2023}). But in order to accurately model the formation of a specific planet one must account for the natal disc chemistry of the planet's birth environment. The initial chemical composition of the protoplanetary disc can be assumed to be the same as that of its host star (e.g. \citealt{Huhn2023}), as they both formed from the same cloud of gas and dust. So the detailed stellar abundance measurements of \citet{Teske2019}, which include 18 of the host stars of the planets from the \citet{Thorngren2016} dataset, allow us to model the growth and evolution of these planets in discs with chemical compositions matched to their birth environment. This is a major improvement on previous models that assume solar-scaled abundances (e.g. \citealt{SchneiderBitsch2021a, Bitsch2023}) and can help explain the observations of \citet{Thorngren2016}.

Atmospheric characterization has entered a golden era thanks to instruments like JWST \citep{Bean2023, Taylor2023}. Transmission spectroscopy is now capable of detecting molecules such as $\mathrm{H_2O}$, CO$_2$, CH$_4$, and sulfur compounds in the atmospheres of transiting giants (e.g.\ WASP‑39b; \citealt{Lueber2024, Powell2024}), and retrieving their C/O ratios and metallicities (e.g. \citealt{Kreidberg2014, August2023, EvansSoma2025}). These atmospheric abundances hold vital clues about the conditions under which planets formed, including their initial positions within the disc, migration pathways, and the timing and nature of material accretion, offering a new route to validate or refute planet formation scenarios \citep{Madhusudhan2017, Booth2017, SchneiderBitsch2021a, Molliere2022, Bitsch2022, Penzlin2024, Ohno2025}.

We use the \texttt{Chemcomp} code \citep{SchneiderBitsch2021a, SchneiderBitsch2021b} to model the growth and migration of the planets within the \citet{Thorngren2016} sample for which stellar abundances are available \citep{Teske2019}. We aim to reproduce the heavy element content of the giant planets within this sample to derive their formation location from their heavy element content.

\section{Methodology}
\label{sec: Methodology}

\subsection{General methods}
The \texttt{Chemcomp} planet formation code \citep{SchneiderBitsch2021a, SchneiderBitsch2021b} is a numerical model designed to simulate the formation of giant planets within protoplanetary discs. Its operation principle involves simulating key processes such as gas and pebble evolution within the disc, planet formation, and planetary migration.

Initially, the model defines a protoplanetary disc with specific initial parameters such as mass ($M_0$), radius ($R_0$), initial metallicity ($\epsilon_0$) and chemical composition. The disc structure, described by the gas surface density $\Sigma_{\text{gas}}$ and temperature $T$, is a power-law profile determined by stellar heating and viscous accretion (e.g. \citet{Lynden-Bell1974, Bell1997}), governed by the viscosity parameter $\alpha$ from \citet{ShakuraSunyaev1973}:
\begin{equation}
\nu = \alpha \frac{c_s^2}{\Omega_K},
\end{equation}
where $c_s$ is the sound speed and $\Omega_K$ is the Keplerian angular velocity.

The isothermal sound speed, $c_s$, is related to the disk midplane 
temperature $T_{\mathrm{mid}}$ through  
\begin{equation}
    c_s = \sqrt{\frac{k_B T_{\mathrm{mid}}}{\mu m_p}},
\end{equation}
where $\mu$ is the mean molecular weight and $m_p$ the proton mass.  
Variations in $\mu$ can occur if the gas is enriched with vapor.

We use a stellar luminosity of $1L_{\odot}$ throughout as this roughly corresponds to the stellar luminosity of stars of age 1.5-2.0Myr \citep{Baraffe2015}. Viscous heating dominates the temperature profile in the inner disc while stellar irradiation only plays a large role in the outer regions \citep{Bitsch2013}. We focus in our study on the inner disc regions, where the heavy element enrichment in the gas phase is largest (e.g. \citealt{Bitsch2023}), indicating that a change in the stellar luminosity would not influence our results.

The viscosity $\nu$ directly influences how quickly material moves radially inward through the disc and onto the star, a process known as viscous accretion. This radial transport is described mathematically by the viscous disc equation \citep{Pringle1981,Armitage2013}  for the gas surface density $\Sigma_{\text{gas}}$:

\begin{equation} \frac{\partial \Sigma_{\text{gas, y}}}{\partial t} - \frac{3}{r}\frac{\partial}{\partial r}\left[r^{1/2}\frac{\partial}{\partial r}\left(\nu\Sigma_{\text{gas, y}}r^{1/2}\right)\right] = \dot{\Sigma}_Y,
\label{eq: Viscous Ev}
\end{equation}

where $\dot{\Sigma}_Y$ is a source term for a given chemical species Y, described below.

Dust grains within the disc grow into pebbles via coagulation, where we follow the approach of \citet{Birnstiel2012}. The inward radial drift speed ($u_Z$) is determined by their Stokes number (St):
\begin{equation}
\text{St} = \frac{\pi}{2}\frac{a\rho_\bullet}{\Sigma_{\text{gas}}},
\end{equation}
with pebble radius $a$ and density $\rho_\bullet$. These pebbles drift towards the star and evaporate upon crossing evaporation lines, enriching the gas in heavy elements like water (H$_2$O) and carbon-bearing species. This enhances the availability of these elements for gas accreting planets forming in these regions, as they accrete the vapor via the gas.

We model the evaporation and condensation of material by including a corresponding term in the viscous evolution equation (Equation \ref{eq: Viscous Ev}). We assume that the inward drifting pebbles would evaporate within $10^{-3}$AU.

Planets in \texttt{Chemcomp} first grow by accreting pebbles, calculated using pebble accretion rates ($\dot{M}_{\text{peb}}$)\citep{JohansenLambrechts2017}, until they reach a critical mass called the pebble isolation mass \citep{Lambrechts2014, Bitsch2018, Ataiee2018} at which point the planet's gravity is strong enough to alter the flow of gas and pebbles around it and opens a gap that blocks inward drifting pebbles. Once the planet reaches the pebble isolation mass, defined as:
\begin{equation}
M_{\text{iso}} \approx 25 \left(\frac{H/r}{0.05}\right)^3 M_{\oplus},
\label{eq:PebIso}
\end{equation}
with $H/r$ being the disc's aspect ratio, pebble accretion ceases, and gas accretion ($\dot{M}_{\text{gas}}$) commences.

Planets migrate first in the type-I regime, where we follow the prescription of \citealt{Paardekooper2011}. We additionally include effects from dynamical torques \citep{Paardekooper2014} as well as from the heating torque \citep{BenitezLlambay2015} that act to slow down the inward migration in the type-I migration regime and could, in some cases, even promote outward migration. Once the planets start to open gaps, their migration rate slows down and they migrate in the slower type-II migration regime with the viscous rate. 

\subsection{Chemistry}
We use \texttt{Chemcomp} \citep{SchneiderBitsch2021a} to simulate the growth and migration of planetary embryos with the goal of matching the heavy element contents of the giant planets measured by \citet{Thorngren2016} In addition, we use the detailed stellar abundances of \citet{Teske2019} as a constraint for our chemical model. This allows the growth of planets to be simulated in discs with chemical compositions matched to their birth environments. We simulate the growth only of the planets heavier than $0.5M_J$ for which host star abundances were available from \citet{Teske2019}. We focus on these massive planets because their compositions are dominated by gaseous envelopes, making their enrichment more difficult to explain using standard models. In contrast, lower-mass planets ($M \ll M_J$) can often match their observed heavy element mass through large solid cores alone (the pebble isolation mass can easily be 10-20$M_\oplus$ alone, see \citealt{Bitsch2018}). Therefore, the most massive planets offer a more stringent test of planet formation theories. The stellar chemical abundances also allow the initial dust-to-gas ratio (DTG) in each disc to be calculated according to the following equation from \citep{Andama2024}:

\begin{equation}
    \text{DTG} = \sum_{\text{X}} \left( \text{X}/{\text{H}} \cdot \mu_{\text{X}} \times 10^{[ \text{X}/\text{H} ]} \right)
\end{equation}

Where X/H is the abundance (by number) of element X (elements used for DTG calculation are C, O, Mg, Si, S and Fe) in a solar-like disc \citep{Asplund2009}, $\mu_\text{X}$ is the atomic mass of element X, and [X/H] is the logarithmic abundance of element X relative to hydrogen, measured in units of dex, from \citet{Teske2019}. The elements included in our simulations are H, He, C, O, Mg, Si, S and Fe. Although sulfur is not included in the abundance measurements of \citet{Teske2019} we can assume that it scales in the same way as silicon \citep{Chen2002}. The initial elemental abundances used in the simulations of each disc along with the dust-to-gas ratios and stellar masses are shown in Table \ref{tab: Stellar Abundances, DTG and Mass} of Appendix \ref{Appendix A}.

Measurements based on solar photospheric and meteoritic CI chondrite abundances suggest that only about 10\% of carbon is found in refractory form, with the remaining 90\% in volatile molecules \citep{Lodders2003}. Cometary data suggest a slightly higher refractory fraction of around 20\% \citep{Altwegg2020}, while interstellar medium (ISM) observations point to much higher values, with up to 60\% of carbon in refractory grains \citep{Bergin2015}. In this paper simulations are run with a refractory carbon fraction of 60\%, consistent with ISM values. This is also consistent with previous studies on giant planet formation and disc chemistry (e.g.  \citealt{SchneiderBitsch2021b, Mah2023}). We do not include the option of refractory carbon burning (e.g. \citealt{Houge2025}) in our model, as this effect does not influence the source of the heavy element content of the planets in our model: the majority of the heavy elements are accreted with the gas exterior to the carbon burning front during planet growth. Furthermore it does not influence the overall message and idea of our work: constraining the origin of the planet via their total heavy element contents.

\subsection{Model parameters}
\begin{figure}
    \centering
    \begin{subfigure}{0.45\textwidth}
        \includegraphics[width=\linewidth]{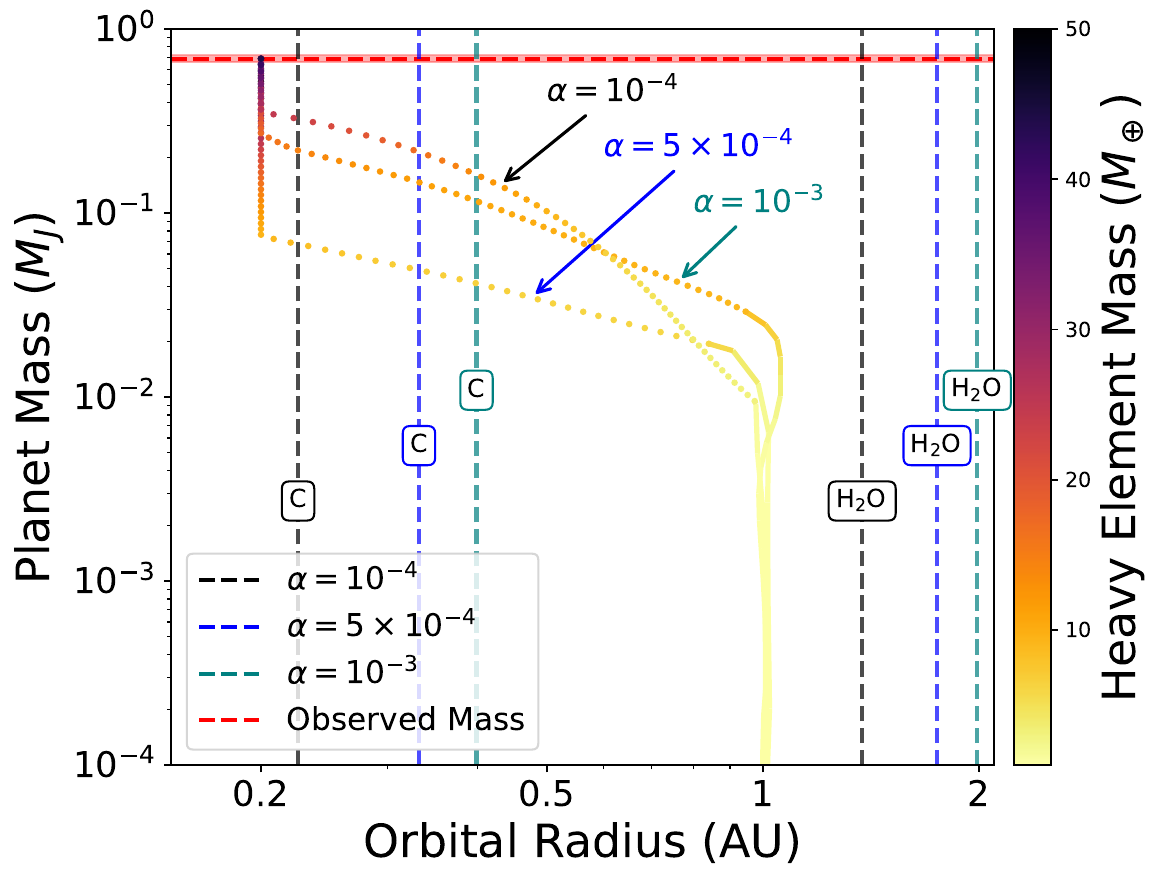}
    \end{subfigure}
    \hfill
    \begin{subfigure}{0.45\textwidth}
        \includegraphics[width=\linewidth]{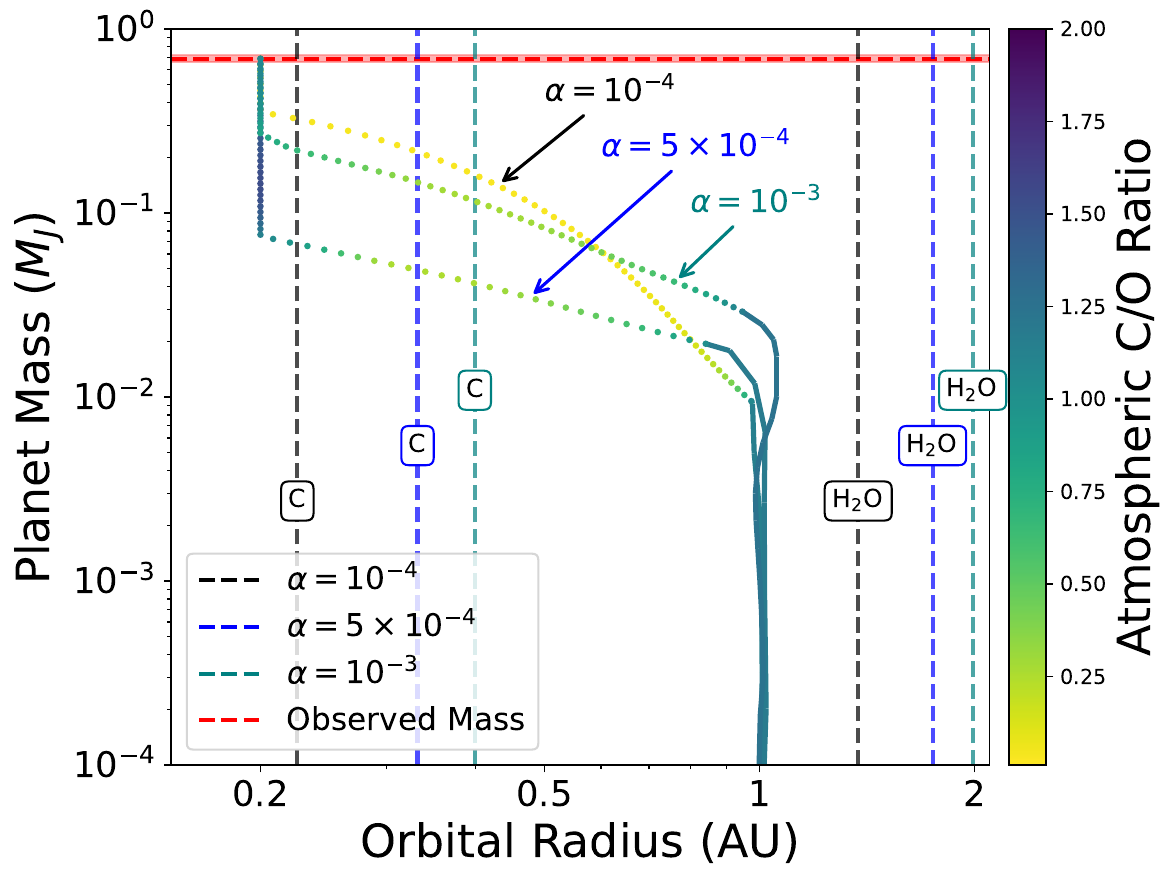}
    \end{subfigure}
    \hfill
    \begin{subfigure}{0.45\textwidth}
        \includegraphics[width=\linewidth]{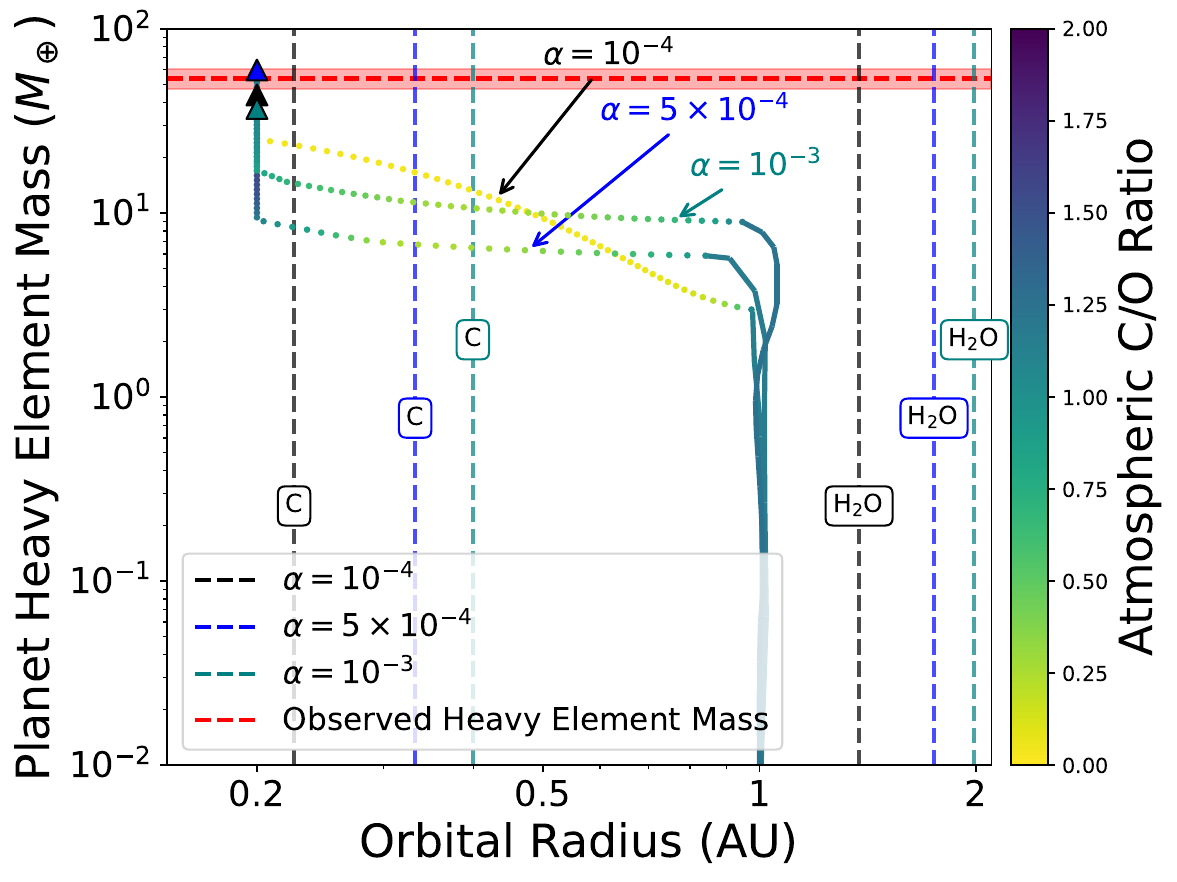}
    \end{subfigure}
    \caption{Growth tracks showing the orbital evolution of a planet simulated in the disc of WASP-84 for 3 different disc viscosities: $\alpha=10^{-4}$, $\alpha=5\times10^{-4}$ and $\alpha=10^{-3}$. The simulations shown are for a planet which originates at an orbital radius of 1AU and migrates inwards. The top plot shows the mass evolution of the planet with the planet's heavy element mass plotted on the colourbar, The middle plot shows the mass evolution of the planet with the planet's atmospheric Carbon to Oxygen ratio plotted on the colourbar and the bottom plot shows the heavy element mass evolution of the planet with the planet's atmospheric Carbon to Oxygen ratio plotted on the colourbar. The locations of the C and H$_2$O evaporation fronts are shown for the 3 different viscosities. The horizontal red dashed line and shaded region shows the observed mass (top and middle) and the observed heavy element mass (bottom) and their uncertainties. The coloured triangles in the bottom plot show the final simulated heavy element mass for each viscosity. Where the growth tracks are plotted as solid lines indicates that the planet is undergoing pebble accretion while dotted lines indicate gas accretion.}
    \label{fig: ViscosityGrowthTracks}
\end{figure}

Simulations are run for viscosity values of $\alpha = 1\times10^{-4}$, $\alpha = 5\times10^{-4}$ and $\alpha = 1\times10^{-3}$. Figure \ref{fig: ViscosityGrowthTracks} shows how a planet's growth and composition varies with different disc viscosities. The timing of a planet's heavy element accretion is heavily dependent on the viscosity of the disc. The early stage of planet growth is the pebble accretion phase where growth is determined by the pebble flux and pebble sizes. In the lower viscosity settings ($\alpha = 1\times10^{-4}$, $\alpha = 5\times10^{-4}$) these are both larger than for high viscosities, and so the core grows slightly faster in these settings. However, once the pebble isolation mass is reached, the accretion rate is limited by the gas accretion rate, which is set by the viscous accretion rate of the disc. There a higher viscosity ($\alpha = 1\times10^{-3}$) automatically results in a faster accretion rate ($\dot{M}_{\rm disc} \propto \nu$). As the majority of the giant planet is composed of gas, this sets the growth time of the planet. All simulations are run with a disc mass of 0.128$M_\odot$ and a disc radius of 137AU \citep{SchneiderBitsch2021a}, a large disc mass is chosen to grow the giant planets efficiently following the idea of \citet{Savvidou2023}. The grain fragmentation velocity used in the simulations is 5 ms$^{-1}$ and planetary embryos are initially placed at 0.1Myr with an initial embryo mass of 0.005$M_\oplus$.

To compare with observations, the output from each simulation is evaluated at the timestep when the simulated planet reaches the observed mass of its corresponding planet from the dataset. While we adopt a fiducial disc lifetime of 3 Myr (this is extended for low viscosity), this method accounts for the fact that protoplanetary discs may disperse earlier in reality (\citealt{Mamajek2009}). For low viscosities ($\alpha=1\times10^{-4}$) long disc lifetimes were required to allow for the formation of the most massive planets for planetary embryos originating interior to the water-ice line due to the slow transport of solid material to the inner disc. Kepler-419b, for example, which has a mass of $\sim2.5M_J$ would have taken $\sim$17Myr to grow to its observed mass if it originated at 0.3AU in a disc of $\alpha=1\times10^{-4}$. HD80606b ($3.94M_J$) and Kepler-432b ($5.84M_J$) would have taken $\sim$12Myr for the same initial scenario. It should be noted that for this initial scenario the planets form with low heavy element masses. For Kepler-419b and HD80606b the final simulated heavy element masses for an initial embryo position of 0.3AU are below the uncertainty range from \citet{Thorngren2016} and so this formation scenario can be dismissed. The observed heavy element mass of Kepler-432b is not well constrained (the uncertainty range extends down to 0.1M$_\oplus$ and so it is very difficult to make a prediction on the planet's formation location in the first place) so while the final simulated heavy element mass for this scenario falls within the uncertainty range we can still dismiss this formation scenario due to the unrealistically long disc lifetime needed to explain its formation. Observations show a spread in disc dispersal timescales, which can be attributed to a range of factors including stellar mass, environment, and photoevaporation rates \citep{Gudel2007, Owens2011, Haworth2018, Picogna2021, Pfalzner2022, Haworth2023}. In particular, variation in high-energy radiation (e.g. X-rays) from young stars can drive differing photoevaporation rates, leading to significant diversity in the lifetimes of protoplanetary discs. Once the planet has reached its desired mass, we assume that the disc dissipates. This is modeled by an exponential decay of the disc's surface density within a short period of time. While we do not model the exact photoevaporative profile of the disc, we note that the planetary composition is unaltered within the two approaches (Sheehan et al., in prep.).

We do not compare the final semi-major axes of the simulations as scattering events after the dispersal of the disc could have moved the planet into its observed orbit \citep{Chatterjee2008, BeaugeNesvorny2012, Paardekooper2018, Bitsch2020}. As scattering events require multiple growing objects it is important to note that the innermost planets grow fastest and are thus only minimally influenced in respect to their composition if outer planets grow afterwards \citep{Eberlein2024}.

In the following section we discuss how varying the initial position of planetary embryos in the disc allows us to directly match the simulated heavy element masses to observations and the predictions that these simulations make about the composition of each planet's atmosphere.

\begin{figure*}[h!]
    \centering
    \includegraphics[width=\linewidth]{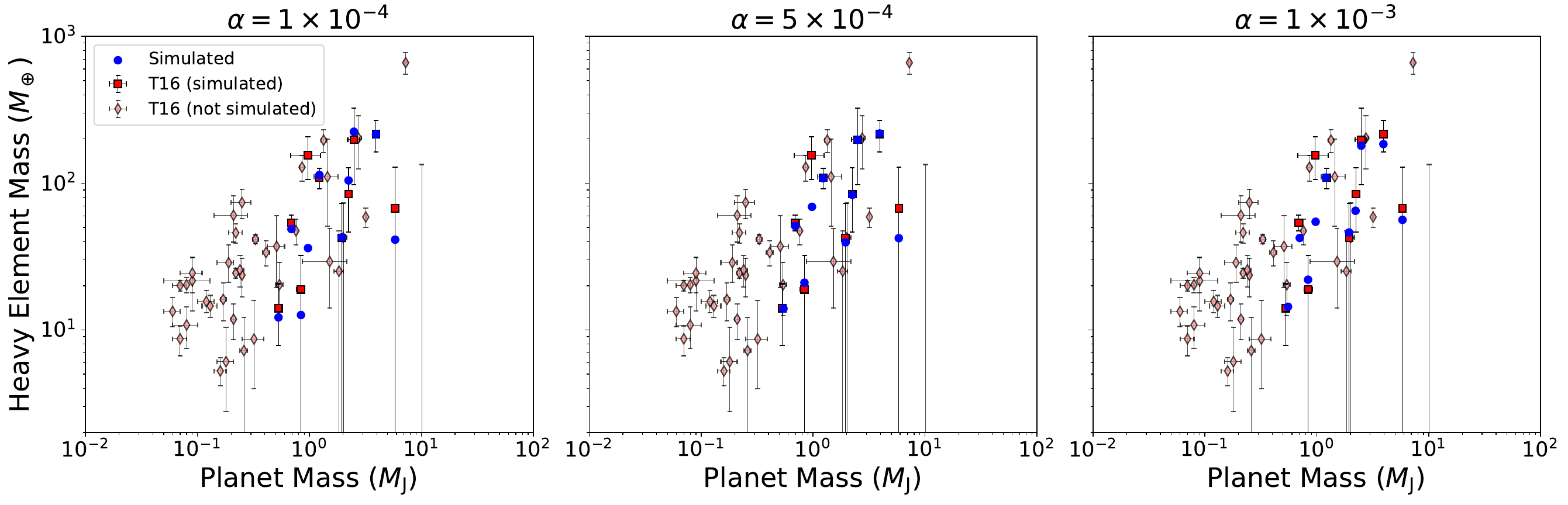}
    \caption{Mass vs Heavy Element Mass relation for the 10 planets simulated. Blue points show simulations where the initial formation location matched the observed heavy element mass most closely. Data from \citet{Thorngren2016} is shown in red (T16). Planets which are simulated in this work are plotted as red squares while those which are not simulated in this work, either because they are less massive than 0.5$M_J$ or because the stellar chemical abundances of their host star are not included in \citet{Teske2019}, are plotted as pale red diamonds.}
    \label{fig:MvsMzMatched}
\end{figure*}

\section{Formation of heavy element enriched planets}

\subsection{Results}
\label{sec:Results}

In this study, we successfully reproduced the observed heavy element masses for 9 out of the 10 most massive planets from \citet{Thorngren2016}, for which stellar abundances observations were available \citep{Teske2019}, for viscosities of $\alpha = 1\times10^{-4}$ and $\alpha = 5\times10^{-4}$ and 8 out of 10 for $\alpha = 1\times10^{-3}$. The simulations could not reproduce the heavy element content of Kepler-539b for any of the viscosity values tested. Kepler-539b is a planet with mass $M = 0.97\,M_J$ and heavy element mass $M_Z = 155\,M_{\oplus}$, corresponding to an unusually high heavy element fraction of 49\%. This level of enrichment makes Kepler-539b much denser than the other planets considered, suggesting that additional processes, such as atmospheric stripping or giant impacts, may have removed a substantial portion of its hydrogen/helium envelope without significantly reducing its core mass \citep{Louden2017, Denman2020}. These mechanisms are not currently accounted for in \texttt{Chemcomp}. 

We match the heavy element content of the giant planets by simulating the growth of planets with different initial positions of the planetary embryo in order to test which initial position would lead to the planet accreting the observed amount of heavy elements once it reached its observed mass. These results are presented in Figure \ref{fig:MvsMzMatched}. For the remainder of the main body of this work we focus on the results from the $\alpha=1\times10^{-4}$ simulations due to the unrealistically rapid planet formation and disc dispersal required to match the heavy element observations for the $\alpha=5\times10^{-4}$ and $\alpha=1\times10^{-3}$ cases. The disc lifetime for a given system was inferred from the timestep at which the simulated planet reached the observed mass and heavy element mass. For the $\alpha=1\times10^{-4}$ discs the inferred lifetimes ranged from 1-10Myr with the majority of planets forming within 4Myr, this is in agreement with the observed distribution of protoplanetary disc lifetimes \citep{Mamajek2009, Pfalzner2022}. For $\alpha=5\times10^{-4}$ discs the inferred lifetimes ranged from 0.2-3Myr with the majority of planets forming within 1Myr and for $\alpha=1\times10^{-3}$ all discs were inferred to disperse within 1Myr. The lifetimes for viscosities larger than $\alpha=1\times10^{-4}$ are not in line with observations. See Appendix \ref{Appendix B} for further justification and an overview of the results for the higher viscosity simulations. Table \ref{tab:formation_regions} explicitly shows the inferred lifetimes of the discs of 5 of the planets simulated which had well constrained heavy element masses for the $\alpha=1\times10^{-4}$ case. In order to illustrate a general trend of how the final planetary heavy element mass and C/O ratio vary with initial position and to obtain possible formation regions where the planet's final heavy element mass matches observations we linearly interpolate the simulation results between adjacent data points.

\begin{figure*}[t]
\centering

\begin{subfigure}[t]{0.49\linewidth}
    \centering
    \includegraphics[width=\linewidth]{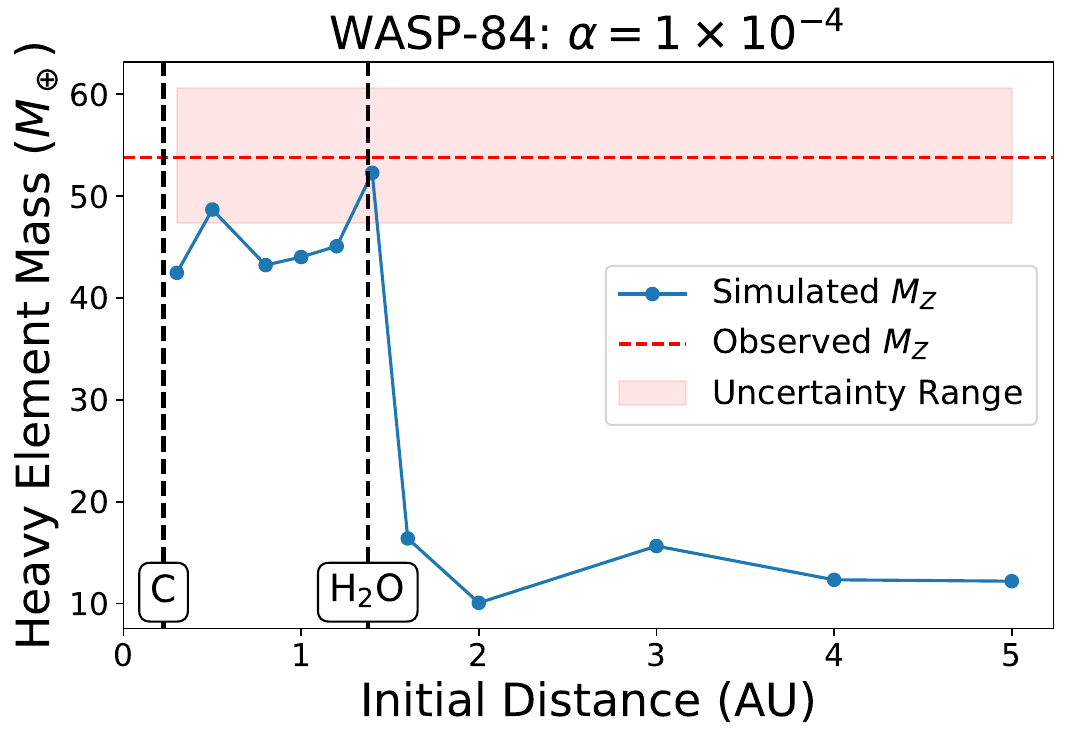}

\end{subfigure}
\hfill
\begin{subfigure}[t]{0.49\linewidth}
    \centering
    \includegraphics[width=\linewidth]{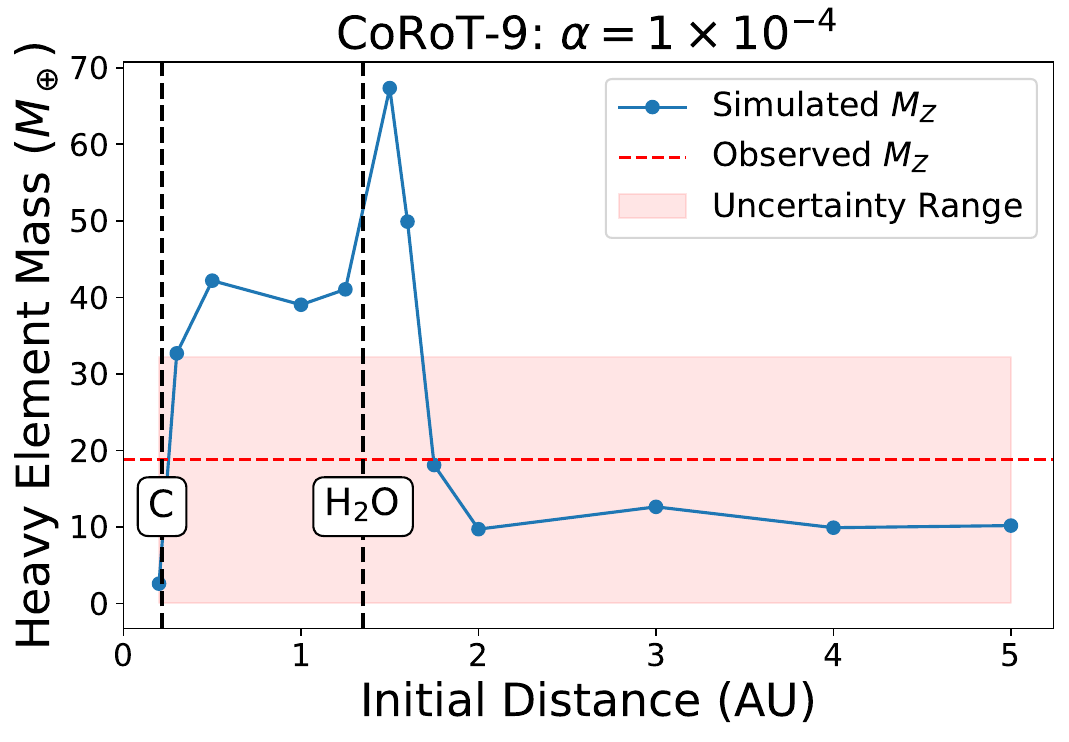}

\end{subfigure}

\vspace{0.5cm}

\begin{subfigure}[t]{0.49\linewidth}
    \centering
    \includegraphics[width=\linewidth]{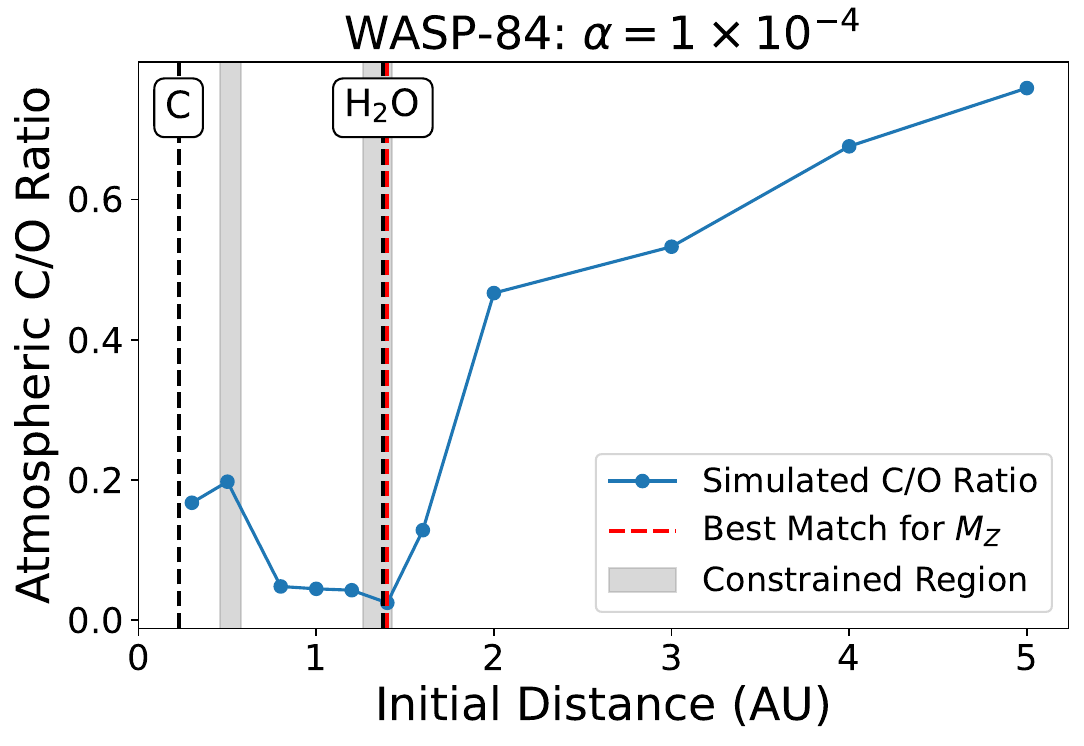}

\end{subfigure}
\hfill
\begin{subfigure}[t]{0.49\linewidth}
    \centering
    \includegraphics[width=\linewidth]{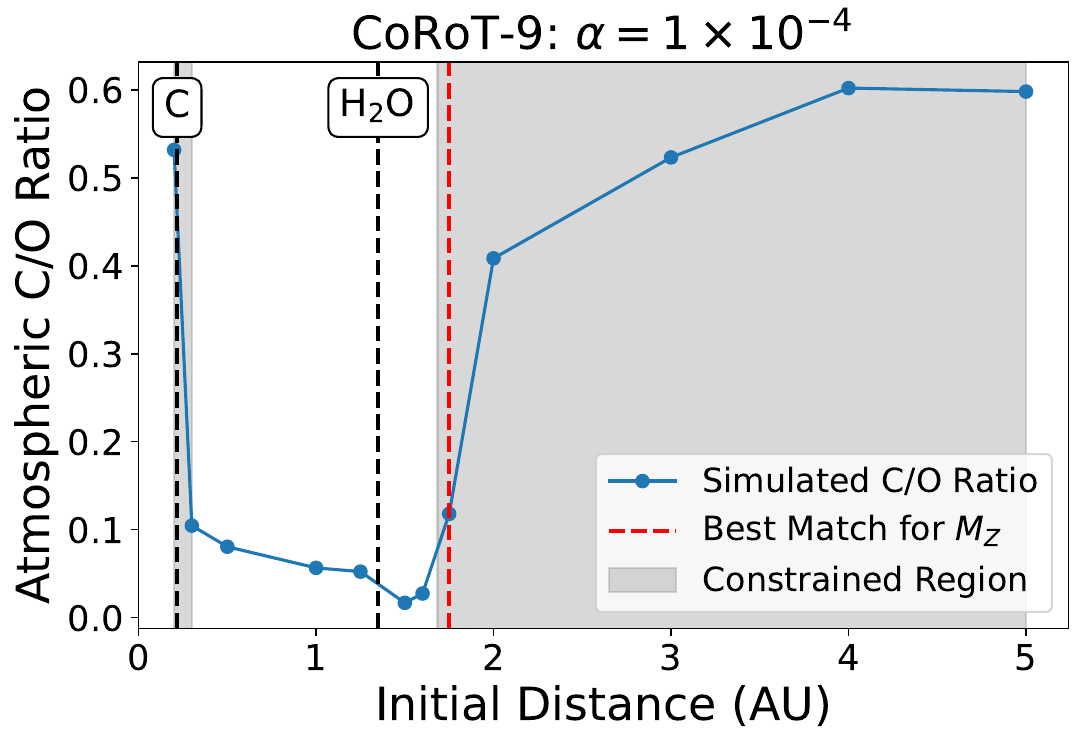}

\end{subfigure}

\caption{Top row: Heavy element masses of the planets simulated in the discs of WASP-84 (left) and CoRoT-9 (right) as a function of the initial position of the planetary embryo. The observed heavy element mass is shown as a dashed red line, with its uncertainty indicated by a red shaded region. Bottom row: Predicted atmospheric C/O number ratios of the same planets as a function of initial position. The red dashed line shows the best-fit formation location based on the heavy element mass, and the grey region indicates the constrained range of possible formation locations. H$_2$O and C evaporation fronts are shown as dashed black lines in all panels. All simulations use $\alpha = 1\times10^{-4}$.}
\label{fig:combined_Mz_CO}
\end{figure*}

The results of our simulations can provide insight into which regions of the protoplanetary disc a planet is most likely to have originated from based solely off its heavy element content and the chemical composition of the disc. This is illustrated in Figure \ref{fig:combined_Mz_CO} for WASP-84b (which has a relatively precise $M_Z$) and CoRoT-9b (which has an imprecise $M_Z$) which shows the importance of precise observational constraints on the heavy element content for this method. A tight constraint on the heavy element mass leads to a more precise prediction on the origin location of the planet. \texttt{Chemcomp} tracks the chemical composition of the growing planet, allowing us to use the inferred origin position to predict properties of the planet's atmosphere. This is illustrated in Figure \ref{fig:combined_Mz_CO} for the Carbon-to-Oxygen ratio of the atmospheres of WASP-84b and CoRoT-9b.

\begin{figure}
    \centering
    \includegraphics[width=0.49\textwidth]{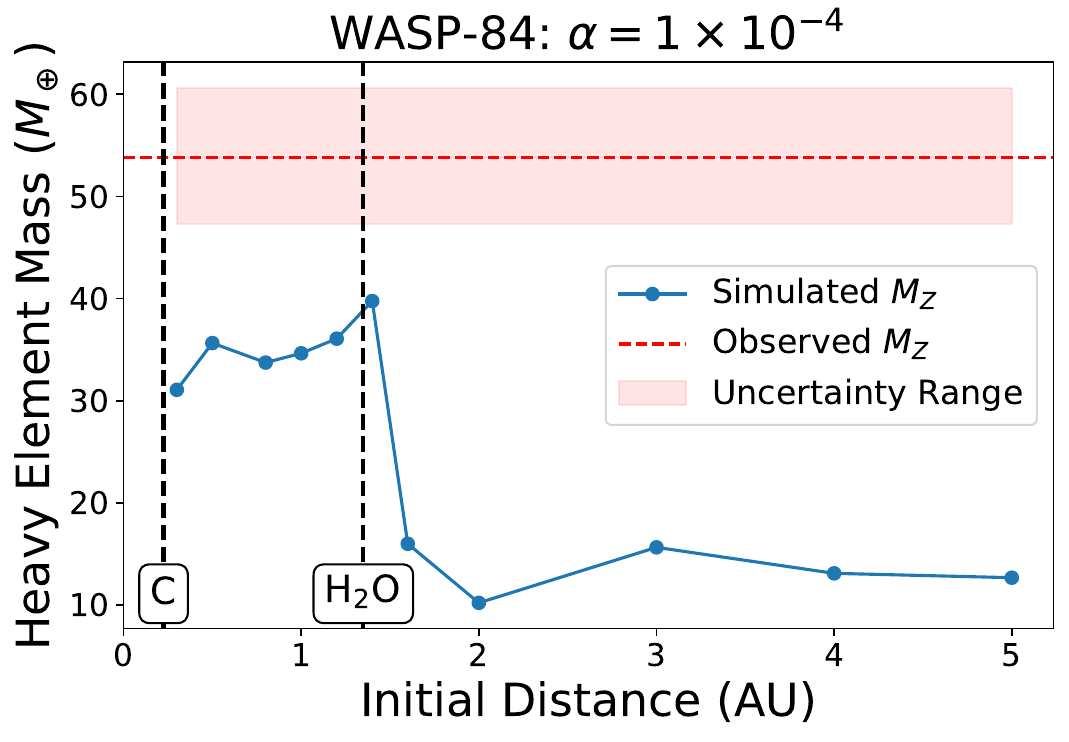}
    \caption{Heavy element masses of the planet simulated in the disc of WASP-84 for varying initial positions of the planetary embryo in a disc of $\alpha=1\times10^{-4}$ with a refractory carbon content of 20\%. The simulated heavy element mass of the planet does not fall within the uncertainty range of observations for any initial position.}
    \label{fig: C20 WASP-84 Mzvsa_p}
\end{figure}

These results show that tighter observational constraints on heavy element mass can significantly improve our ability to infer formation pathways and atmospheric compositions. Our ability to match the observations of \citet{Thorngren2016} depends strongly on the refractory carbon fraction, Figure \ref{fig: C20 WASP-84 Mzvsa_p} shows that the formation of planets with large heavy element contents, such as WASP-84b, cannot be explained if we assume the lower bound of the refractory carbon fraction of 20\%. A reduced refractory carbon fraction increases the abundances of CO and CO$_2$, while decreasing that of H$_2$O as a larger proportion of the available oxygen is contained in CO and CO$_2$. Because CO and CO$_2$ sublimate exterior to the water ice line, beyond the region where our planets must form to reach the observed enrichment, they cannot contribute significantly to the enrichment of the inner disc gas during early growth. The simultaneous depletion of H$_2$O further limits the available heavy elements for accretion in the inner disc, compounding this effect. Hence we do not explore the results of simulations with lower refractory carbon fractions in this work.

\begin{figure}[h]
    \centering
    \includegraphics[width=\linewidth]{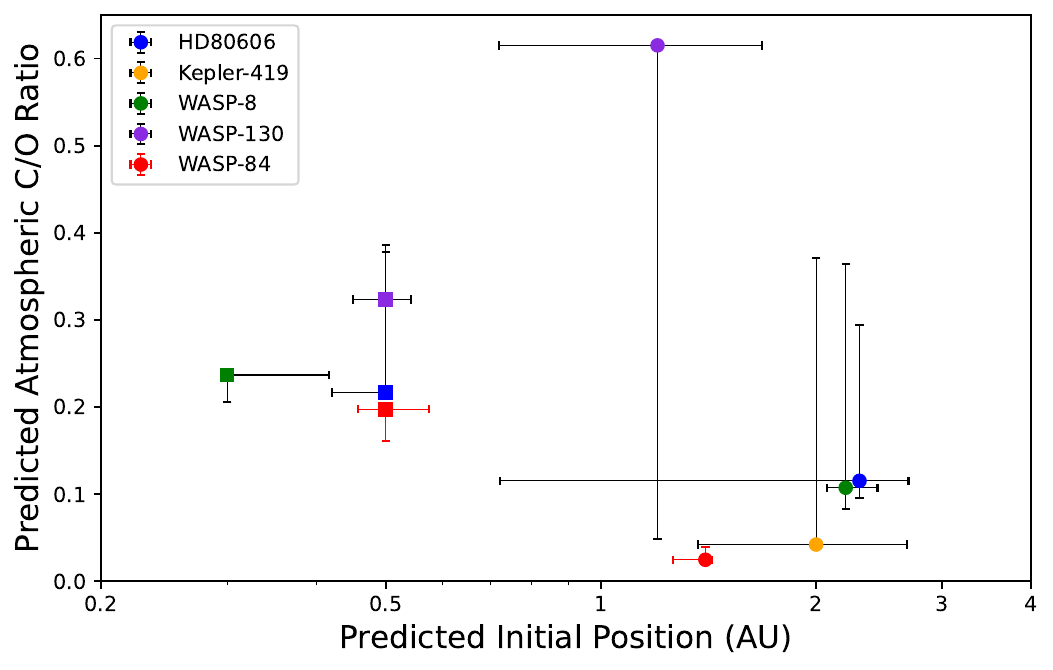}
    \caption{Predicted initial positions which simulate the observed heavy element masses and the corresponding predicted C/O ratios for the 5 planets simulated which had good constraints on the heavy element content. All simulations use $\alpha = 1\times10^{-4}$. Square points represent the 'inner' formation regions and circles represent the 'outer' formation regions. The points are placed at the initial position which led to the simulations most closely matching the observed heavy element mass. The error bars on the x axis represent the width of the constrained possible formation regions of each planet while the error bars on the y axis show the range in the predicted C/O ratio within each region. WASP-84b is the focus of this work due to its potential observability and is marked with red points and error bars.}
    \label{fig:COpredictions}
\end{figure}

Figure \ref{fig:COpredictions} shows atmospheric C/O predictions for planets forming in discs with $\alpha = 1\times10^{-4}$ which had good constraints on their heavy element content, allowing for tighter constraints on the formation region and thus the C/O ratio. These constrained formations regions and C/O ratios, along with the inferred final semi-major axes and disc lifetimes, are shown in Table \ref{tab:formation_regions}. Planets with uncertainties in their heavy element contents which extended down no lower than to 5\% of the total planet mass were identified as being well constrained. Conversely, planets with heavy element measurements with error bars that extended down to almost zero, such as CoRoT-9b, were identified as being poorly constrained as their heavy element content could be explained solely by their core mass. For such planets we can only exclude formation in a small region in the inner disc for which the planet would have accreted too much heavy elements to match observations. In contrast, planets with well constrained heavy element masses often have 2 distinct possible formation regions, one in the very inner disc (interior to the water-ice line) and one just beyond the water-ice line. However, as the planet grows it migrates and crosses the water ice line, allowing it to accrete large fractions of heavy elements (see Appendix \ref{Appendix B}). These 2 formation regions can be seen for WASP-84b and CoRoT-9b in Figure \ref{fig:combined_Mz_CO}. For both planets 2 peaks in the simulated heavy element mass exist, however the peak in simulated heavy element mass at the H$_2$O evaporation front is much more pronounced compared to the peak before the C evaporation front for CoRoT-9b than it is for WASP-84b. This can be explained by the abundances in Table \ref{tab: Stellar Abundances, DTG and Mass}. The disc of WASP-84 has a super-solar carbon abundance and a sub-solar oxygen abundance leading to a greater heavy element enrichment of the gas at the C evaporation front and a lesser enrichment at the H$_2$O evaporation front when compared to a solar disc, because the larger carbon fraction allows more C-grains as well as CO and CO$_2$, which take away oxygen from forming H$_2$O. The disc of CoRoT-9b has sub-solar carbon and oxygen abundances, with a slightly larger O abundance, leading to a stronger peak in simulated heavy element mass at the H$_2$O front compared to the C front. The proportion of total planetary mass made up of heavy elements is observed to be approximately 25\% for WASP-84b and 7\% for CoRoT-9b. These observations can only be explained if WASP-84b formed initially either close to the H$_2$O front or just beyond the C evaporation front, maximising the simulated metal enrichment. CoRoT-9b's low enrichment can be explained by a large range of formation positions. 

There was no inner formation region found for Kepler-419b, it can only accrete the observed amount of heavy elements if it begins to form exterior to the water-ice line due to the timing of material accretion. If a planet grows rapidly in the very inner disc at early times its atmosphere becomes less metal enriched than if it had formed slightly further out, beyond the water-ice line, as icy pebbles have not yet drifted inwards and enriched the gas phase. In the inner disc, interior to the water ice line, refractory solids (silicates and metals) are abundant enough to allow the formation of planet cores of a few Earth masses that can then continue to accrete volatile rich gas. Especially close to the carbon evaporation line enough vapour exists to push the total heavy element content of the planet growing there into the regime of WASP-84b.

\subsection{Impact of model assumptions}
\subsubsection{Fragmentation velocity}

The grain fragmentation velocity sets the size of the pebbles in the inner disc region, where their size is dominated by fragmentation rather than by drift, which dominates the pebble sizes in the outer disc regions. We use here a fixed fragmentation velocity of 5m/s. Using a lower fragmentation velocity of 1m/s would reduce the grain size by a factor of 25 \citep{Birnstiel2012}. As such, the pebbles would drift slower, however, they still remain very large and can enrich the inner disc with vapour to large values \citep{Bitsch2023}. The enrichment with heavy elements in the inner disc regions is determined by two effects: the inward drifting velocity of the pebbles that set how fast the inner disc gets enriched (between 0.5-1.0 Myr), while the disc's viscosity sets how long the inner disc can stay enriched, because the viscosity determines the inward velocity of the gas.  In particular, the low viscosity that allows us to match the heavy element content of WASP-84b, results in a long enrichment of the vapour phase of the inner disc independently of the grain fragmentation velocity (see \citealt{Bitsch2023}). We thus think that the effects of the grain fragmentation velocity are minimal for our here proposed mechanism to constrain the formation location of giant planets via their total heavy element content.

Interior to the water ice line, the pebbles lose their icy component, resulting in smaller pebble sizes with Stokes numbers of the order of $10^{-3}$ to $10^{-2}$ for low fragmentation velocities of 1m/s and up to a factor of 100 larger if the fragmentation velocity is increased to 10m/s (e.g. \citealt{Bitsch2023}). The low Stokes numbers are sometimes associated with difficulties to build cores via pebble accretion in these regions (e.g. \citealt{BatyginMorbidelli2023}), however the formation of planetesimals via the streaming instabilties at these low Stokes numbers is also difficult (e.g. \citealt{LiYoudin2021}). If the planets grow via pebble accretion, their final mass is set by the pebble isolation mass, which scales strongly with the disc's aspect ratio (e.g. \citealt{Lambrechts2014, Bitsch2018}). However, at low Stokes numbers, the disc might be hotter due to the increased opacity of the large grains compared to the small grains, resulting in aspect ratios high enough to form cores capable of accreting gas (e.g. \citealt{SavvidouBitsch2021}), if they can grow efficiently enough. In reality, probably a mixture of growth mechanism (pebble accretion, planetesimal accretion, collision between growing embryos) needs to operate to allow cores to grow massive enough to accrete a gaseous envelope and transition into gas giants. This process is not modelled here in detail and also generally not fully understood. However, our study shows that if cores form in these inner disc regions, they will accrete a large fraction of heavy elements via the gas, allowing them to match the large heavy element content of the observed giant planets, independently of how the cores reached masses large enough to allow efficient gas accretion.

\subsubsection{Initial embryo formation time and mass}

Planetary embryos below the so-called pebble transition mass accrete in the slow Bondi regime, while planets above the pebble transition mass accrete in the Hill regime \citep{LambrechtsJohansen2012, JohansenLambrechts2017}. If we were to start our embryos at a lower mass, they would need a longer time to grow. Essentially, it means we could mimic this effect by putting embryos of the same mass at later times. However, if planetary embryos are inserted at later times, most of the pebbles have already drifted inwards, preventing their growth to become gas giants \citep{Savvidou2023}. In order to form giant planets, we thus require an early planetary embryo insertion time.

\section{Discussion and conclusions}
\label{sec: Discussion}

Formation location and migration history influence not only the total heavy element content but also the atmospheric carbon-to-oxygen (C/O) ratio \citep{Oberg2011, Madhusudhan2014,  Turrini2021, Bitsch2022, Penzlin2024}. As planets migrate through chemically distinct regions, they accrete gas with varying C and O abundances. For example, accretion inside the water ice line results in oxygen-rich gas, lowering the C/O ratio, while accretion beyond this line yields higher C/O ratios.

Because \texttt{Chemcomp} tracks elemental abundances, we can also predict the C/O ratio for each formation location, for each simulation and initial position we have a corresponding C/O ratio. Combining these C/O values with the constrained formation regions provides predicted ranges for each planet’s atmospheric composition (the minimum to maximum C/O within the constrained formation region). Figure \ref{fig:combined_Mz_CO} shows how both the final heavy element mass and the atmospheric C/O ratio of the planet changes with initial position. For both planets there are 2 constrained regions where the linear interpolation of the simulated heavy element mass is within the uncertainty range. So we have 2 predicted ranges for the C/O ratio of each planet corresponding to the range in simulated C/O values within each region.

The range of predicted C/O depends on how well the planet’s initial position is constrained. For planets with poorly measured heavy element masses, this range is wider. Figure \ref{fig:combined_Mz_CO} illustrates this for CoRoT-9b, whose poorly constrained heavy element mass results in a large range of possible formation locations and therefore a broad predicted C/O range.

While we demonstrated how the carbon to oxygen ratio of the planetary atmosphere could be predicted for planets whose formation location is constrained via their total heavy element content, we can also make predictions about the other chemical elements used in our simulations.  Appendix \ref{Appendix D} shows the O/H and C/H predictions for WASP-84b and Corot-9b. In Table \ref{tab: WASP-84 abu} we summarise these findings and also include predictions of Mg, Si, S and Fe within the atmosphere of WASP-84b for both its inner and outer formation regions, giving additional constraints for observations. The predicted O/H enrichment compared to the host star is dominated by water which is accreted by the growing planet interior to the H$_2$O evaporation front, while the large C/H enrichment in the inner formation region is due to Carbon grains which can be accreted into the atmosphere when the planet moves interior to the C evaporation front. The large sulfur abundance compared to other refractories is due to the evaporation of H$_2$S, which is entrapped in water ice \citep{Santos2025}, at the H$_2$O evaporation front, enriching the gas in the inner disc with sulfur. H$_2$S is also enhanced compared to a solar composition due to the slightly super solar S abundance of WASP-84. 

\begin{table*}[h!]
\caption{WASP-84b predicted atmospheric abundances for the inner and outer formation regions.}
\centering
\begin{tabular}{lcccccccc}
\hline
\textbf{Region} & \textbf{Element} & \multicolumn{2}{c}{\textbf{Minimum}} & \multicolumn{2}{c}{\textbf{Maximum}} & \multicolumn{2}{c}{\textbf{Best Match}} \\
 & & Number Ratio & Normalised Ratio & Number Ratio & Normalised Ratio & Number Ratio & Normalised Ratio \\
\hline
\multirow{7}{*}{Inner} 
& C/O  & 0.161 & 0.252 & 0.197 & 0.308 & 0.197 & 0.308 \\
& C/H  & $3.05\times10^{-3}$ & 10.45 & $3.77\times10^{-3}$ & 12.89 & $3.77\times10^{-3}$ & 12.89 \\
& O/H  & $1.85\times10^{-2}$ & 40.50 & $1.91\times10^{-2}$ & 41.72 & $1.91\times10^{-2}$ & 41.72 \\
& Mg/H & $1.06\times10^{-5}$ & 0.26 & $1.1\times10^{-5}$ & 0.27 & $1.1\times10^{-5}$ & 0.27 \\
& Si/H & $9.74\times10^{-6}$ & 0.26 & $1.01\times10^{-5}$ & 0.27 & $1.01\times10^{-5}$ & 0.27 \\
& S/H  & $1.41\times10^{-4}$ & 9.20 & $1.47\times10^{-4}$ & 9.54 & $1.47\times10^{-4}$ & 9.54 \\
& Fe/H & $9.46\times10^{-6}$ & 0.27 & $9.72\times10^{-6}$ & 0.27 & $9.72\times10^{-6}$ & 0.27 \\
\hline
\multirow{7}{*}{Outer} 
& C/O  & 0.025 & 0.039 & 0.039 & 0.061 & 0.025 & 0.039 \\
& C/H  & $5.76\times10^{-4}$ & 1.97 & $7.42\times10^{-4}$ & 2.94 & $5.76\times10^{-4}$ & 1.98 \\
& O/H  & $2.05\times10^{-2}$ & 44.84 & $2.34\times10^{-2}$ & 51.18 & $2.34\times10^{-2}$ & 51.18 \\
& Mg/H & $6.08\times10^{-6}$ & 0.15 & $7.9\times10^{-6}$ & 0.19 & $6.08\times10^{-6}$ & 0.15 \\
& Si/H & $5.58\times10^{-6}$ & 0.15 & $7.25\times10^{-6}$ & 0.19 & $5.58\times10^{-6}$ & 0.15 \\
& S/H  & $1.57\times10^{-4}$ & 10.21 & $1.82\times10^{-4}$ & 11.85 & $1.82\times10^{-4}$ & 11.85 \\
& Fe/H & $5.25\times10^{-6}$ & 0.15 & $6.82\times10^{-6}$ & 0.19 & $5.25\times10^{-6}$ & 0.15 \\
\hline
\end{tabular}
\tablefoot{For each element, Number Ratio is the absolute number ratio, and Normalised Ratio is the value normalised to the host star’s abundance. The minimum and maximum values correspond to the range within the constrained region, and the best match is the ratio at the data point which most closely matches the heavy element content of WASP-84b.}
\label{tab: WASP-84 abu}
\end{table*}

By calculating the transmission spectroscopy metric (TSM), see \citet{Kempton2018}, of the planets in the sample with well constrained heavy element contents WASP-84b was identified as a prime target for atmospheric characterization (this is described in Appendix \ref{Appendix C}). This would provide a means of testing the concept put forward in this paper. 

The recent observations of WASP-121b's atmosphere revealed a very enriched atmosphere in Carbon, Oxygen and Silicon \citep{EvansSoma2025}, which can be explained by accretion of gas enriched in these elements. Our simulations for WASP-84b show indeed similar predicted enrichments for Oxygen and Carbon, but not for Silicon. This could be attributed to the fact that WASP-121b accreted most of its mass even closer to the host star, where also the rock forming species evaporate and are accreted via the gas. WASP-121b orbits with a period of 1.3 days. While such an ultra-short-period may result from tidal interaction with the central star, followed by planet-planet scattering rather than pure disc-driven migration, migration may still have brought the planet close to the star before such events occurred. In this case the heavy element accretion could be aided by the mechanisms that we investigate in this work.  While other mechanisms such as the late accretion of dust, comets, or planetesimals may contribute to the observed enrichment (e.g. \citealt{Arras2022, Morbidelli2023, Bitsch2023, Seligman2022, Shibata2024}), planetesimal accretion alone is unlikely to explain the high heavy element content. In fact, efficient planetesimal formation generally reduces the heavy-element fraction of forming giant planets, as planetesimal accretion is less efficient than the accretion of vapour-enriched gas \citep{Danti2023}. Putting this together, the large total heavy element content of giant planets implies a large enrichment of the atmosphere, especially in carbon and oxygen, explained within our model. This offers a testable prediction of the pebble drift and evaporation model.

Our model is also consistent with the new results of WASP-80b's composition \citep{AcuneAguirre2025} in respect to the heavy element content, C/O and core mass. The method described in our work requires stellar elemental abundances as an input for our formation model which are not available for WASP-80 and so we can not derive a definite formation location of the planet. To confirm the general consistency of our model with the results of \citet{AcuneAguirre2025} we estimated the stellar abundances of WASP-80 according to the scaling relation of \citet{BitschBattistini2020} where only the [Fe/H] of WASP-80 \citep{Terrien2015} is required as an input. \citet{Bardet2025} presents atmospheric C/O ratio retrievals for ten hot Jupiters based on JWST transmission spectra. Their findings show that the retrieved C/O values mostly lie between $\sim$0.1 and 0.6, with median values concentrated in the range 0.2–0.4. These observationally derived values fall within the same broad range as the predicted atmospheric C/O ratios shown in Figure \ref{fig:COpredictions} of this work. Although the planets analysed by \citet{Bardet2025} are different from those modelled here, the overlap in C/O ranges highlights the plausibility of our predictions and the need for follow up observations to verify the efficacy of the methods put forward in this paper for determining the birthplaces of exoplanets based off their heavy element content.

\begin{acknowledgements}
    Barry O'Donovan acknowledges the support of the European Research Council (ERC) Starting Grant (ExoPEA: 01164652). We
thank the referee for the comments that helped to improve this
work. 
\end{acknowledgements}

\bibliographystyle{aa} 
\bibliography{references}

\begin{appendix}
\onecolumn
\section{Stellar abundances}
\label{Appendix A}
Table \ref{tab: Stellar Abundances, DTG and Mass} shows all the stars from the \citet{Teske2019} dataset along with their stellar abundances, our calculated DTG ratios, stellar masses and the total planetary mass and planetary heavy element mass of the corresponding planet from the \citet{Thorngren2016} dataset. As explained in Section \ref{sec: Methodology} we simulate the formation only of those planets which are above 0.5$M_J$ and so Table \ref{tab: Stellar Abundances, DTG and Mass} includes only the stars from the \citet{Teske2019} dataset which hosted planets above this threshold mass.

\renewcommand{\arraystretch}{1.3}
\begin{longtable}{l c c c c c c c c c}
\caption{Stellar abundances \citep{Teske2019}, computed dust-to-gas ratios (DTG), stellar masses, planetary mass and planetary heavy element mass for each disc simulated in this work.}
\label{tab: Stellar Abundances, DTG and Mass} \\[3pt]
\toprule
\textbf{Star} & \textbf{[C/H]} & \textbf{[O/H]} & \textbf{[Mg/H]} & \textbf{[Si/H]} &
\textbf{[Fe/H]} & \textbf{DTG} & \textbf{Mass ($M_\odot$)} &
\textbf{Planet Mass ($M_J$)} & \textbf{M$_Z$ ($M_\oplus$)} \\
\midrule
\endfirsthead

\caption{continued. Stellar abundances \citep{Teske2019}, computed dust-to-gas ratios (DTG), and stellar masses.}\\[3pt]
\toprule
\textbf{Star} & \textbf{[C/H]} & \textbf{[O/H]} & \textbf{[Mg/H]} & \textbf{[Si/H]} &
\textbf{[Fe/H]} & \textbf{DTG} & \textbf{Mass ($M_\odot$)} &
\textbf{Planet Mass ($M_J$)} & \textbf{M$_Z$ ($M_\oplus$)} \\
\midrule
\endhead

\midrule
\multicolumn{10}{r}{\textit{Continued on next page}}
\endfoot

\bottomrule
\endlastfoot

HD80606    &  0.270 &  0.200 & 0.328 & 0.328 &  0.28 & $2.669 \times 10^{-2}$ & 1.05 & $3.94 \pm 0.11$ & $215.69^{+52.63}_{-52.23}$ \\
Kepler-432 & -0.103 & -0.085 & 0.005 & 0.070 &  0.02 & $1.338 \times 10^{-2}$ & 1.35 & $5.84 \pm 0.05$ & $67.39^{+60.95}_{-67.29}$ \\
Kepler-419 &  0.131 &  0.131 & 0.010 & 0.117 &  0.03 & $1.959 \times 10^{-2}$ & 1.39 & $2.50 \pm 0.30$ & $198.16^{+126.80}_{-100.72}$ \\
WASP-8     &  0.149 &  0.153 & 0.250 & 0.273 &  0.29 & $2.335 \times 10^{-2}$ & 0.99 & $2.24^{+0.08}_{-0.09}$ & $84.34^{+43.33}_{-38.00}$ \\
HAT-P-15   &  0.182 &  0.118 & 0.212 & 0.265 &  0.27 & $2.251 \times 10^{-2}$ & 1.01 & $1.95 \pm 0.07$ & $42.45^{+30.67}_{-42.35}$ \\
WASP-130   &  0.274 &  0.179 & 0.269 & 0.322 &  0.29 & $2.593 \times 10^{-2}$ & 1.04 & $1.23 \pm 0.04$ & $108.83^{+17.58}_{-17.35}$ \\
Kepler-539 & -0.188 & -0.047 &-0.040 &-0.099 & -0.10 & $1.247 \times 10^{-2}$ & 1.05 & $0.97 \pm 0.29$ & $154.94^{+51.90}_{-48.76}$ \\
CoRoT-9    & -0.111 & -0.041 & 0.007 &-0.028 & -0.02 & $1.355 \times 10^{-2}$ & 0.96 & $0.84 \pm 0.07$ & $18.87^{+13.30}_{-18.77}$ \\
WASP-84    &  0.036 & -0.030 & 0.014 & 0.066 &  0.05 & $1.535 \times 10^{-2}$ & 0.84 & $0.69 \pm 0.03$ & $53.82^{+6.79}_{-6.48}$ \\
HAT-P-17   &  0.053 &  0.019 & 0.062 & 0.060 &  0.02 & $1.633 \times 10^{-2}$ & 0.86 & $0.53 \pm 0.02$ & $14.08^{+6.56}_{-6.24}$ \\
\end{longtable}
\renewcommand{\arraystretch}{1.0}

\clearpage

\section{Planet semi major axis evolution and the effects of varying viscosity}
\label{Appendix B}

\begin{figure*}[h!]
    \centering
    \begin{subfigure}[t]{1\linewidth}
    \centering
    \includegraphics[width=\linewidth]{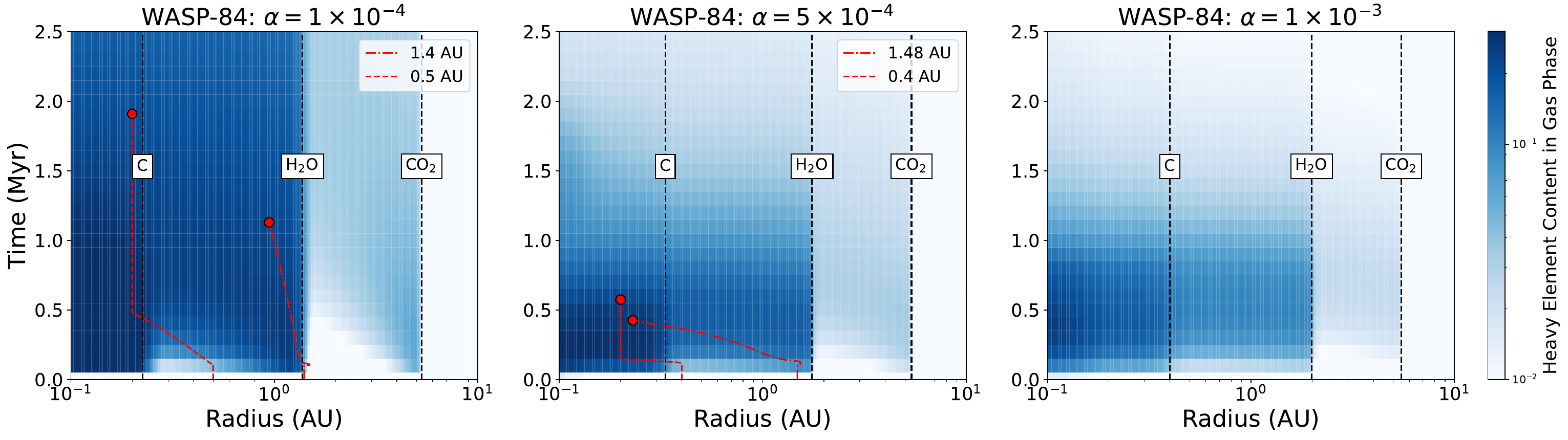}
\end{subfigure}
\vspace{0.5cm}
\begin{subfigure}[t]{1\linewidth}
    \centering
    \includegraphics[width=\linewidth]{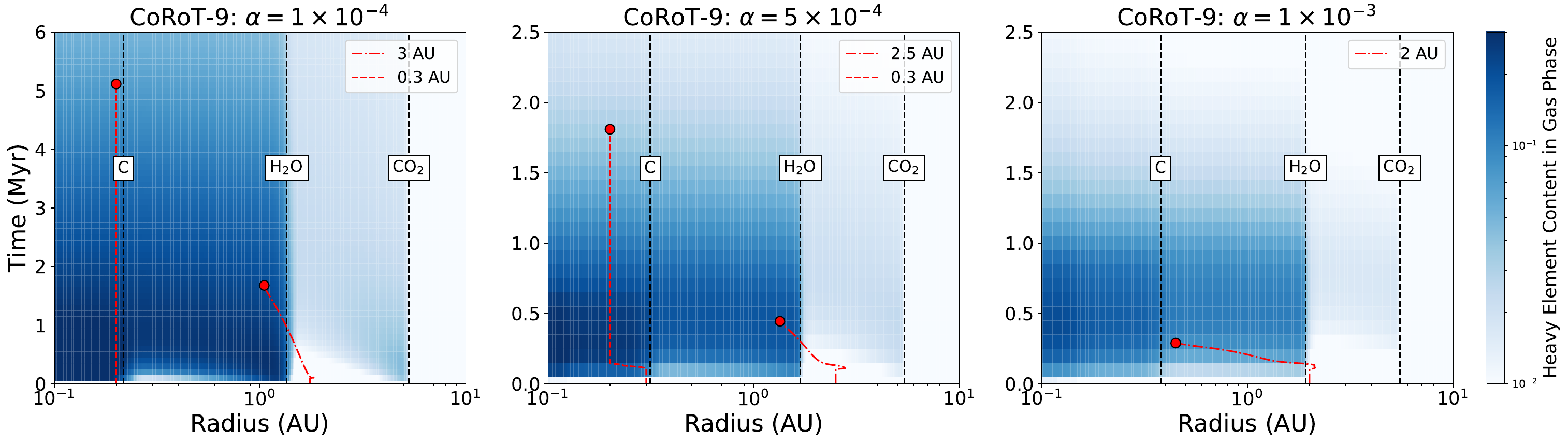}
\end{subfigure}
    \caption{Semi major axis evolutions of planets forming in the disc of WASP-84 (top) and CoRoT-9 (bottom) for discs of varying viscosity:
$\alpha=1\times10^{-4}$ (left), $\alpha=5\times10^{-4}$ (middle) and $\alpha=1\times10^{-3}$ (right). Red circles indicate the time and position at which the planet reached its observed mass for each simulation. All evolutions plotted are for simulations which matched the observed heavy element masses of WASP-84b and CoRoT-9b. In the disc of WASP-84, for $\alpha=1\times10^{-3}$, the observed heavy element mass could not be matched by simulations for any initial position.}
    \label{fig: Growth Tracks}
\end{figure*}

\begin{figure*}[t]
\centering
\begin{subfigure}[t]{0.49\linewidth}
    \centering
    \includegraphics[width=\linewidth]{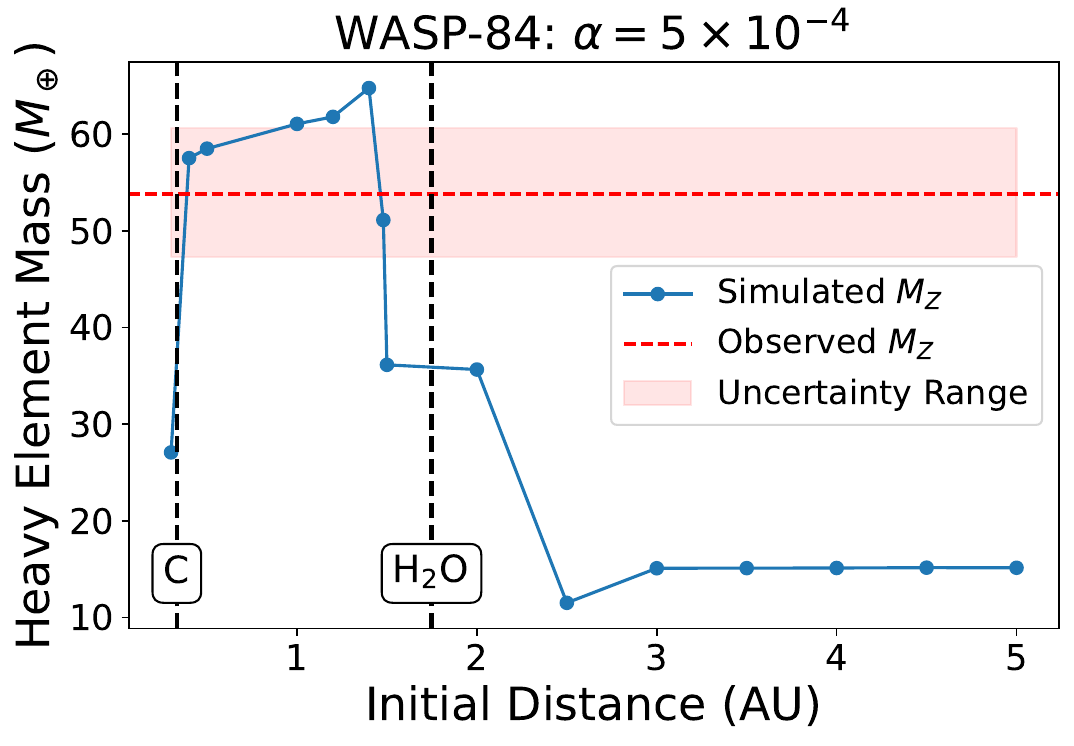}
\end{subfigure}
\hfill
\begin{subfigure}[t]{0.49\linewidth}
    \centering
    \includegraphics[width=\linewidth]{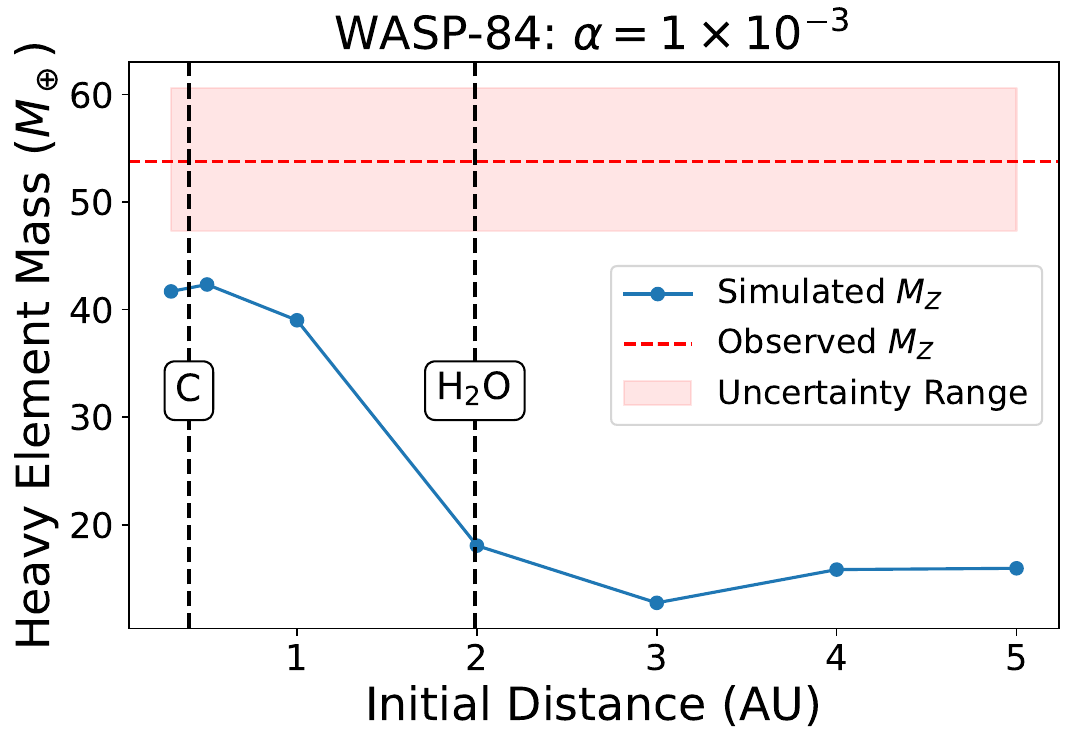}
\end{subfigure}
    \caption{Heavy element masses of the planet simulated in the disc of WASP-84 for varying initial positions of the planetary embryo in a disc of $\alpha=5\times10^{-4}$ (left) and in a disc of $\alpha=1\times10^{-3}$ (right). The simulated heavy element mass of the planet in the $\alpha=1\times10^{-3}$ disc does not fall within the uncertainty range of observations for any initial position whereas the $\alpha=5\times10^{-4}$ disc has 2 separate possible formation regions, like the $\alpha=1\times10^{-4}$ disc (see Figure \ref{fig:combined_Mz_CO}).}
    \label{fig: WASP-84 alpha1e-3 and 5e-4 Mzvsa_p}
\end{figure*}

\begin{figure*}[t]
\centering
\begin{subfigure}[t]{0.49\linewidth}
    \centering
    \includegraphics[width=\linewidth]{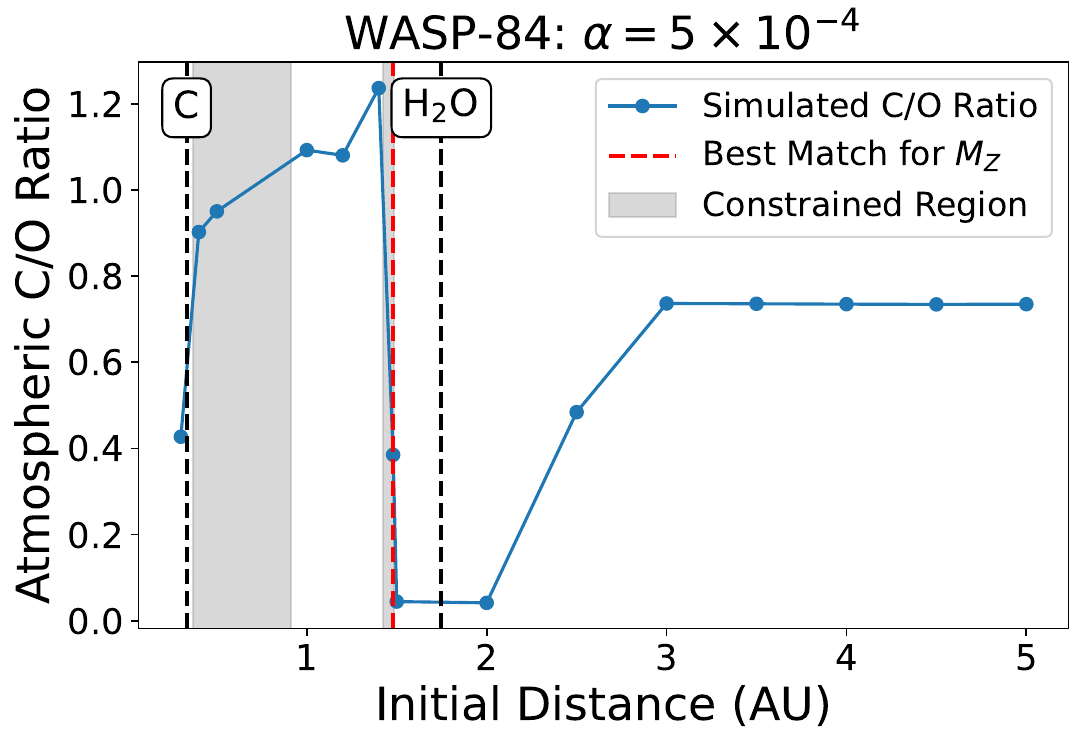}
\end{subfigure}
\hfill
\begin{subfigure}[t]{0.49\linewidth}
    \centering
    \includegraphics[width=\linewidth]{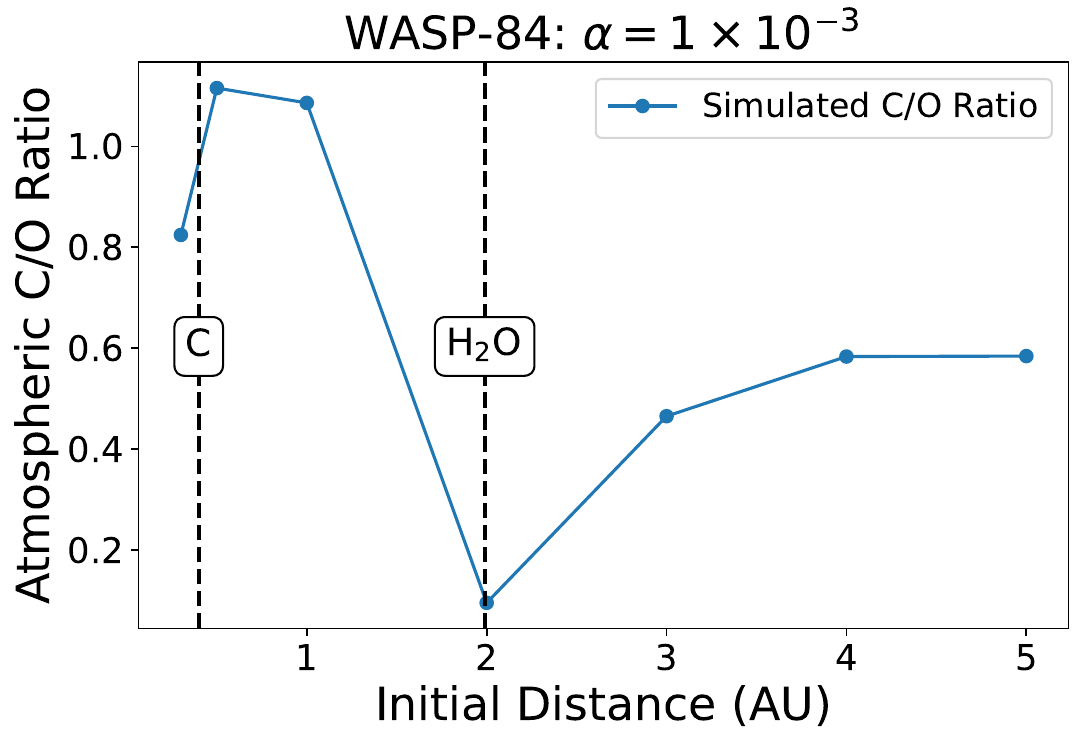}
\end{subfigure}
    \caption{Predicted atmospheric C/O number ratios of the planet simulated in the disc of WASP-84 as a function of initial position for a disc of $\alpha=5\times10^{-4}$ (left) and $\alpha=1\times10^{-3}$ (right). The red dashed line shows the best-fit formation location based on the heavy element mass, and the grey region indicates the constrained range of possible formation locations. H$_2$O and C evaporation fronts are shown as dashed black lines in all panels. No possible formation regions were identified for WASP-84b in a disc of $\alpha=1\times10^{-3}$.}
    \label{fig: WASP-84 alpha1e-3 and 5e-4 C/O}
\end{figure*}

Figure \ref{fig: Growth Tracks} shows the semi major axis evolution of the planets which accreted the observed amount of heavy elements for the 3 different viscosities simulated in the discs of WASP-84 and CoRoT-9. The heavy element content in the gas phase in the protoplanetary disc is plotted as a colourmap, with darker blue indicating more heavy element enrichment. The evaporation fronts of C, H$_2$O and CO$_2$ are also plotted. \texttt{Chemcomp} stops migration planet migration at 0.2AU just before the inner edge of the disc grid as grid cells interior to the planet's position are required to calculate the gradients needed to compute the planet's migration and gap opening. Physically this can be justified by the existence of a migration trap in the inner disc halting the planet's migration (e.g. \citealt{Flock2019}). For some planets and viscosities (e.g. both CoRoT-9b and WASP-84b at $\alpha=5\times10^{-4}$ and $\alpha=1\times10^{-4}$) there are 2 possible formation regions that lead to the planet accreting the observed amount of heavy elements, Figure \ref{fig:combined_Mz_CO} illustrates this for CoRoT-9b and WASP-84b at $\alpha=1\times10^{-4}$ and Figure \ref{fig: WASP-84 alpha1e-3 and 5e-4 Mzvsa_p} does so for $\alpha=5\times10^{-4}$. These tracks demonstrate that although outer planets begin in less metal enriched regions, they migrate into the inner disc closer to the water ice line, where the initial increase with vapour is largest, allowing them to become enriched in heavy elements. An $\alpha=1\times10^{-3}$ disc fails to reproduce the observed heavy element enrichment of WASP-84b for all initial positions, this is illustrated further in Figure \ref{fig: WASP-84 alpha1e-3 and 5e-4 Mzvsa_p}. In general high $\alpha$ discs ($\alpha=5\times10^{-4}$ and $\alpha=1\times10^{-3}$) often result in very rapid growth in the early stages of the disc ($<$ 1Myr). At the same time, the enrichment of the disc with heavy elements is not large enough to explain the heavy element contents of the giant planets. For example, WASP-84b has roughly a heavy element content of 25\% of its mass, but the discs with high viscosity do not reach enrichments of heavy elements larger than 10\%. In addition, we stop the simulations when the planet reaches its observed mass, this would require the disc to disperse in unrealistically short timescales. For this reason we focus on the $\alpha=1\times10^{-4}$ simulations in this paper. However, for completeness the simulated C/O of WASP-84b is plotted as a function of initial position for WASP-84b in discs of $\alpha=5\times10^{-4}$ and $\alpha=1\times10^{-3}$ in Figure \ref{fig: WASP-84 alpha1e-3 and 5e-4 C/O}. It is important to note that there is no constrained region and thus no predicted C/O ratio for the $\alpha=1\times10^{-3}$ case as we are unable to match the observed heavy element mass of WASP-84b at this viscosity.

\renewcommand{\arraystretch}{1.4} 
\begin{table*}[h!]
\caption{Constrained initial planetary embryo semi-major axes, final semi-major axes (i.e. the planet's position when it reaches its observed mass), C/O number ratios and disc lifetimes for an $\alpha = 10^{-4}$ disc.}
\centering
\begin{tabular}{lcccccc}
\hline
\textbf{Planet} & \textbf{Region} & \textbf{Initial $a_p$ (AU)} & \textbf{Final $a_p$ (AU)} & \textbf{C/O (Number Ratio)} & \textbf{Disc Lifetime (Myr)}  \\
\hline
HD80606b   & Inner & 0.5$_{-0.08}$ & 0.2 & 0.22$^{+0.16}$ & 4.48$^{+3.06}$  \\
           & Outer & 2.3$^{+0.39}_{-1.58}$ & 0.39$^{+0.29}_{-0.19}$ &  0.12$^{+0.18}_{-0.02}$ & 5.56$^{+0.16}_{-0.99}$  \\
\hline
Kepler-419b & Inner & - & - & - & -  \\
           & Outer & 2.0$^{+0.68}_{-0.63}$ & 0.53$^{+0.53}_{-0.21}$ & 0.042$^{+0.33}$ & 4.17$^{+7.12}$   \\
\hline
WASP-8b   & Inner & 0.3$^{+0.12}$ & 0.2 & 0.24$_{-0.03}$ & 7.08$_{-2.29}$ \\ 
           & Outer & 2.2$^{+0.24}_{-0.13}$ & 0.88$^{+0.21}_{-0.07}$ & 0.11$^{+0.26}_{-0.02}$ &  3.52$^{+0.06}_{-0.18}$ \\
\hline
WASP-130b  & Inner & 0.5$^{+0.04}_{-0.05}$ & 0.2 & 0.32$^{+0.04}$ & 1.18$^{+0.71}_{-0.03}$  \\
           & Outer & 1.2$^{+0.48}_{-0.48}$ & 0.2$^{+0.85}$ & 0.62$_{-0.57}$ & 3.3$_{-1.55}$  \\
\hline
WASP-84b  & Inner & 0.5$^{+0.07}_{-0.04}$ & 0.2 & 0.2$^{-0.04}$ & 1.91$^{+0.11}$  \\
           & Outer & 1.4$^{+0.03}_{-0.14}$ & 0.94$^{+0.02}_{-0.44}$ & 0.03$^{+0.01}$ & 1.13$^{+0.56}$  \\
\arrayrulecolor{black}\hline
\end{tabular}
\tablefoot{The final semi-major axis is defined as the planet's position when it reaches its observed mass in the simulation. Shown are the values for the simulation which most closely matched the observed heavy element mass with the uncertainties corresponding to the range of these quantities within the constrained formation regions (all quantities are linearly interpolated between each simulation data point). There is no constrained inner formation region for Kepler-419b.}
\label{tab:formation_regions}
\end{table*}
\renewcommand{\arraystretch}{1.0}

\FloatBarrier

\section{TSM calculation}
\label{Appendix C}
 \citet{Kempton2018} present a framework for quantifying how amenable an exoplanet is to atmospheric characterisation. The transmission spectroscopy metric (TSM), which is proportional to the expected transmission spectroscopy signal to noise ratio, is calculated as follows:

\begin{equation}
    \text{TSM} = \text{(scale factor)} \times \frac{R_p^3T_{eq}}{M_p R_*^2} \times 10^{-m_J/5} 
\label{eq:TSM}
\end{equation}

The quantities in Equation \ref{eq:TSM} are defined as follows:

\begin{enumerate} 
    \item $R_p$: the radius of the planet in units of Earth radii,
    
    \item $M_p$: the mass of the planet in units of Earth masses, 

    \item $R_\star$: the radius of the host star in units of solar radii,

    \item $T_{\text{eq}}$: the planet’s equilibrium temperature in Kelvin calculated for zero albedo and full day-night heat redistribution according to
    \begin{equation}
    T_{\text{eq}} = T_\star \sqrt{\frac{R_\star}{a}} \left( \frac{1}{4} \right)^{1/4},
    \label{eq:Teq}
    \end{equation}
    where $T_\star$ is the host star effective temperature in Kelvin, and $a$ is the orbital semimajor axis given in the same units as $R_\star$,

    \item $m_J$: the apparent magnitude of the host star in the J band.
\end{enumerate}

The "scale factor" in Equation \ref{eq:TSM} is a normalisation constant selected to give one-to-one scaling between the analytic transmission metric of \citet{Kempton2018} and the more detailed work of \citet{Louie2018}. \citet{Kempton2018} calculates the scale factor separately for each planet radius bin, their sample uses radial bins up to $R_p < 10R_\oplus$. Since the planets we wish to calculate the TSM for (HD80606b, Kepler-419b, WASP-8b, WASP-130b and WASP-84b) have radii of between $10R_\oplus$ and $12.7R_\oplus$ we use the scale factor of 1.15 calculated for the upper most radial bin in \citet{Kempton2018}.

\begin{table}[htbp]
\caption{Transmission Spectroscopy Metric (TSM) values for selected giant exoplanets.}
\centering
\begin{tabular}{lc}
\hline
\textbf{Planet} & \textbf{TSM} \\
\hline
HD~80606b      &  14.7 \\
Kepler-419b    & 1.4 \\
WASP-8b        & 55.6 \\
WASP-84b      & 115.9 \\
WASP-130b       & 22.5 \\
\hline
\end{tabular}
\tablefoot{TSM values are calculated following \citet{Kempton2018}. Values needed to calculate TSM were obtained from the \href{https://exoplanetarchive.ipac.caltech.edu/}{NASA Exoplanet Archive}.}
\label{tab:tsm}
\end{table}

The TSM values presented in Table~\ref{tab:tsm} provide a means of assessing the relative suitability of the selected planets for atmospheric characterisation via transmission spectroscopy. \citet{Kempton2018} propose a TSM threshold of $\sim$90 for giant planets ($4 < R_p < 10\,R_\oplus$) to identify high-priority targets for JWST/NIRISS observations. While this guideline does not strictly extend to planets with radii exceeding $10\,R_\oplus$, it remains a useful benchmark. Of the planets considered here, WASP-84b stands out with a TSM of 115.9, exceeding the recommended threshold and indicating strong potential for atmospheric follow-up. WASP-8b also shows a moderately high TSM (55.6), suggesting some potential suitability. For completeness, the predicted atmospheric abundances for WASP-8b are shown in Table \ref{tab: WASP-8 abu}. In contrast, HD~80606b, WASP-130b, and Kepler-419b have significantly lower TSM values (14.7, 22.5, and 1.4, respectively), implying low transmission signal-to-noise. Notably, HD~80606b, despite its large mass and radius, is known for its extreme orbital eccentricity \citep{Laughlin2009}, which may complicate atmospheric observations and their interpretation. These results reinforce the utility of the TSM as a prioritisation tool, as demonstrated in the validation work of \citet{Louie2018}, and support the selection of WASP-84b as a strong candidate for atmospheric spectroscopy to test our formation theory.

\begin{table}[h!]
\caption{WASP-8b predicted atmospheric abundances for the inner and outer formation regions.}
\label{tab: WASP-8 abu}
\centering
\begin{tabular}{l c c c c c c c}
\toprule
\textbf{Region} & \textbf{Element} &
\multicolumn{2}{c}{\textbf{Minimum}} &
\multicolumn{2}{c}{\textbf{Maximum}} &
\multicolumn{2}{c}{\textbf{Best Match}} \\
 & &
Number Ratio & Normalised Ratio &
Number Ratio & Normalised Ratio &
Number Ratio & Normalised Ratio \\
\midrule
\multirow{7}{*}{Inner}
& C/O  & 0.205 & 0.377 & 0.237 & 0.436 & 0.237 & 0.436 \\
& C/H  & $1.43\times10^{-3}$ & 3.77 & $3.0\times10^{-3}$ & 7.92 & $1.43\times10^{-3}$ & 3.77 \\
& O/H  & $6.04\times10^{-3}$ & 8.67 & $1.56\times10^{-2}$ & 22.45 & $6.04\times10^{-3}$ & 8.67 \\
& Mg/H & $1.18\times10^{-6}$ & 0.02 & $2.76\times10^{-6}$ & 0.04 & $1.18\times10^{-6}$ & 0.02 \\
& Si/H & $1.01\times10^{-6}$ & 0.02 & $2.37\times10^{-6}$ & 0.04 & $1.01\times10^{-6}$ & 0.02 \\
& S/H  & $4.9\times10^{-5}$ & 1.98 & $2.41\times10^{-4}$ & 9.75 & $4.9\times10^{-5}$ & 1.98 \\
& Fe/H & $1.62\times10^{-5}$ & 0.26 & $1.3\times10^{-4}$ & 2.11 & $1.62\times10^{-5}$ & 0.26 \\
\midrule
\multirow{7}{*}{Outer}
& C/O  & 0.083 & 0.153 & 0.402 & 0.739 & 0.107 & 0.197 \\
& C/H  & $1.31\times10^{-3}$ & 3.45 & $1.43\times10^{-3}$ & 3.78 & $1.34\times10^{-3}$ & 3.55 \\
& O/H  & $3.56\times10^{-3}$ & 5.10 & $1.64\times10^{-2}$ & 23.54 & $1.25\times10^{-2}$ & 17.96 \\
& Mg/H & $3.91\times10^{-6}$ & 0.06 & $4.03\times10^{-6}$ & 0.06 & $4.03\times10^{-6}$ & 0.06 \\
& Si/H & $3.35\times10^{-6}$ & 0.06 & $3.46\times10^{-6}$ & 0.06 & $3.46\times10^{-6}$ & 0.06 \\
& S/H  & $1.3\times10^{-5}$ & 0.53 & $1.19\times10^{-4}$ & 4.81 & $8.58\times10^{-5}$ & 3.47 \\
& Fe/H & $3.4\times10^{-6}$ & 0.06 & $3.51\times10^{-6}$ & 0.06 & $3.51\times10^{-6}$ & 0.06 \\
\bottomrule
\end{tabular}
\end{table}

\FloatBarrier

\section{Oxygen and carbon abundances}
\label{Appendix D}

\begin{figure*}[h!]
\centering
\begin{subfigure}[t]{0.46\linewidth}
    \centering
    \begin{subfigure}[t]{\linewidth}
        \centering
        \includegraphics[width=\linewidth]{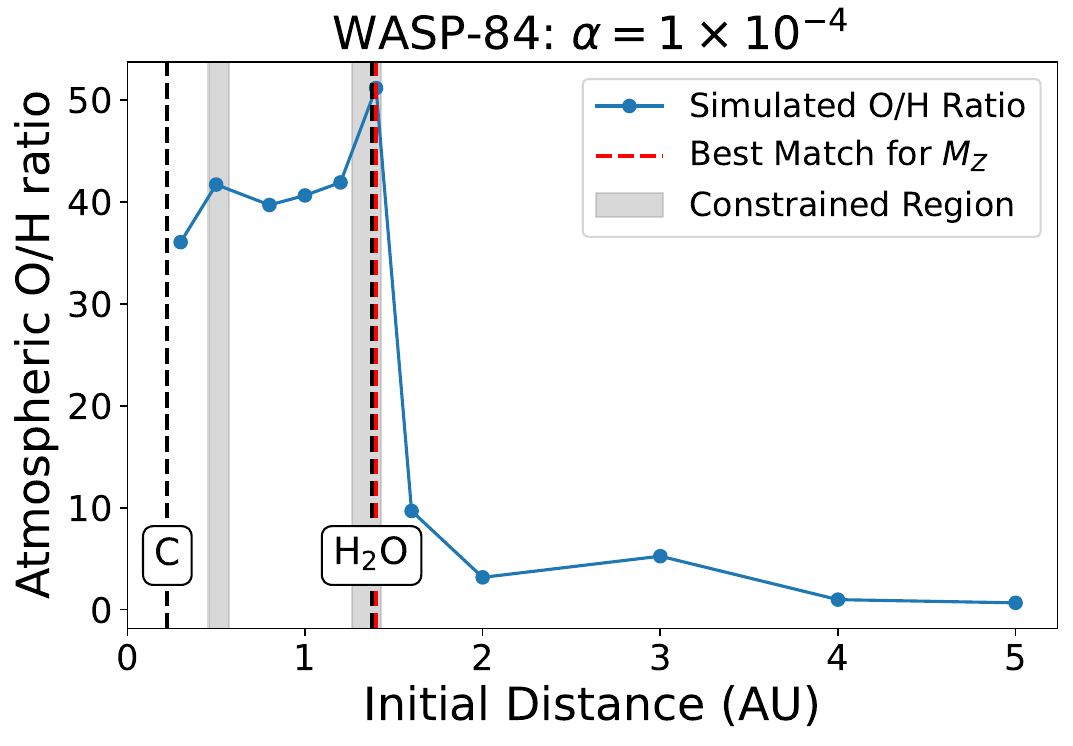}
    \end{subfigure}
    \vspace{0.5cm}
    \begin{subfigure}[t]{\linewidth}
        \centering
        \includegraphics[width=\linewidth]{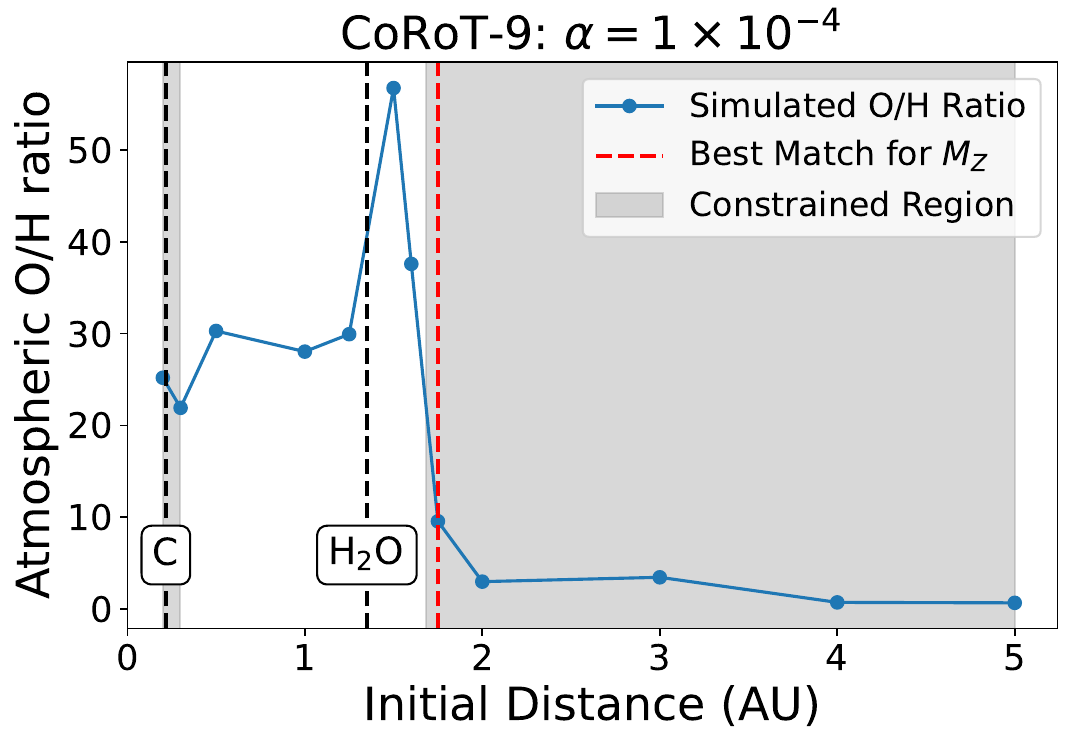}
    \end{subfigure}
\end{subfigure}
\hfill
\begin{subfigure}[t]{0.46\linewidth}
    \centering
    \begin{subfigure}[t]{\linewidth}
        \centering
        \includegraphics[width=\linewidth]{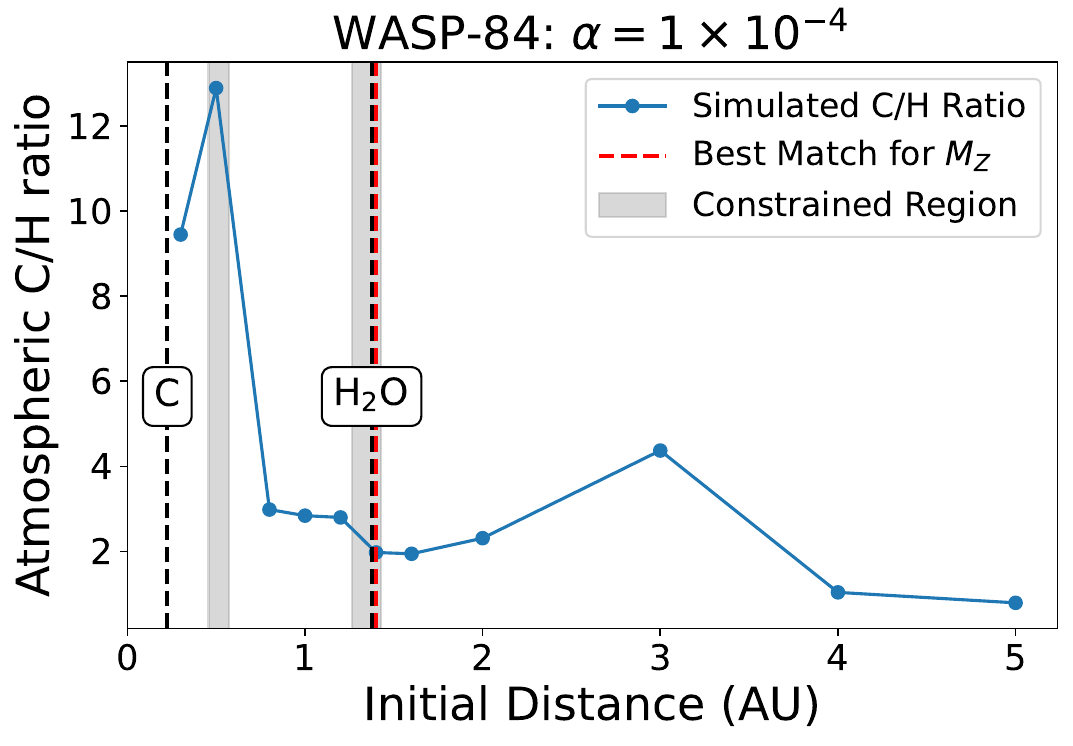}
    \end{subfigure}
    \vspace{0.5cm}
    \begin{subfigure}[t]{\linewidth}
        \centering
        \includegraphics[width=\linewidth]{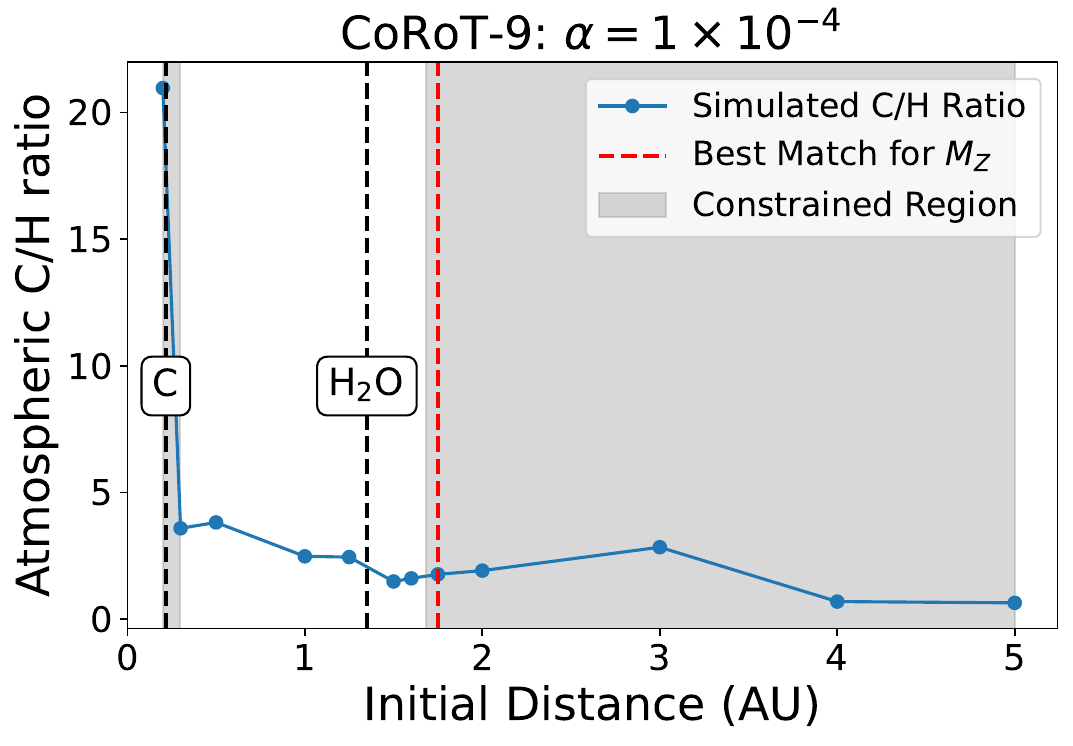}
    \end{subfigure}
\end{subfigure}

\caption{O/H and C/H (relative to host star abundance) of the atmosphere of the planets simulated in the disc of WASP-84 (top) and CoRoT-9 (bottom) for varying initial positions of the planetary embryo in a disc of $\alpha=1\times10^{-4}$. The formation location which resulted in the simulated heavy element mass of the planet matching most closely with the observation is shown as a dashed red line, the constrained regions where the planet could have initially formed are shown as shaded grey regions, and the H$_2$O and C evaporation fronts are shown as dashed black lines.}
\label{fig:OH_CH_vsa_p}
\end{figure*}

Similarly to Figure \ref{fig:combined_Mz_CO}, Figure \ref{fig:OH_CH_vsa_p} shows how the atmospheric O/H and C/H ratios change with the initial position of the planetary embryo. Once again, a good constraint on the heavy element content is essential for obtaining precise predictions of the atmospheric properties of the planet.

\end{appendix}
\end{document}